\documentstyle[aps,eqsecnum,epsfig]{revtex}
\epsfclipon
\tighten
\begin{document}
\title{\bf Spinodal Decomposition and Inflation: Dynamics and
Metric Perturbations}
\author{{\bf D. Cormier$^{(a)}$, R. Holman$^{(b)}$}}
\address
{(a) Institute for Physics, University of Dortmund, D-44221 Dortmund,
Germany \\
(b) Department of Physics, Carnegie Mellon University, Pittsburgh,
PA. 15213, U. S. A.}
\date{December, 1999}
\maketitle

\begin{abstract}
We analyse the dynamics of spinodal decomposition in inflationary cosmology
using the closed time path formalism of out of equilibrium quantum field theory
combined with the non-perturbative Hartree approximation.  In addition to a
general analysis, we compute the detailed evolution of two inflationary models
of particular importance: $\lambda \Phi^4$ new inflation and natural inflation.
We compute the metric fluctuations resulting from inflationary phase
transitions in the slow roll approximation, showing that there exists a regime
for which quantum fluctuations of the inflaton field result in a significant
deviation in the predictions of the spectrum of primordial density
perturbations from standard results. We provide case examples for which a blue
tilt to the power spectrum (i.e. $n_s>1$) results from the evolution of a
single inflaton field, and demonstrate that field fluctuations may result in a
scalar amplitude of fluctuations significantly below standard predictions,
resulting in a slight alleviation of the inflationary fine tuning problem.  We
show explicitly that the metric perturbation spectrum resulting from inflation
depends upon the state at the outset of the inflationary phase.
\end{abstract}

\section{Introduction}
In recent years, there has been a strong interest in the {\it dynamics} of
quantum fields in the early universe.  This interest has led to a better
understanding of a number of processes including the formation of topological
defects during early phase transitions\cite{defects}, the reheating of the
universe after inflation\cite{reheating}, and the dynamics of inflation
itself\cite{inflation}.  In the particular case of inflationary reheating, our
improved understanding has been revolutionary and has significantly reshaped
the subject\cite{preheating}.

The lessons provided by these studies are varied.  One crucial aspect is the
importance of using time-dependent techniques to study processes of the early
universe.  It has been repeatedly shown that classical and one-loop effective
potentials are poorly defined and of little use in dynamical systems; they
should only be used to determine static quantities such as the ground state of
the system\cite{useneq}.  Another common theme is the importance of non-linear
corrections to the linear dynamics.  These corrections have been found to be
quite dramatic in studies of phase transitions and reheating.

Despite these important advances in studies of quantum fields in the early
universe, it is still widely believed that these techniques have little to add
to our understanding of the inflationary phase itself. The belief that the
inflaton follows a classical trajectory determined from the classical effective
potential with only perturbatively small quantum corrections\cite{standard} is
still widely held. While the techniques of stochastic
inflation\cite{stochastic} allow these corrections to add up, they do so in an
incoherent fashion through a repeated summation of one-loop effects without
self-consistent inclusion of higher order corrections\cite{stochcrit}.

Much work has been done to verify that the dynamics of the inflaton field is
predominantly classical.  The existence of a particle horizon and the natural
squeezing of states due to the near exponential redshifting of field modes
justifies a quasi-classical description of the inflaton field up to a
perturbatively small component of field fluctuations with wavelengths inside
the de Sitter horizon\cite{classicalization}.  However, it is important to
emphasize that the validity of a classical evolution of the inflaton field
alone does {\it not} justify the use of the classical effective potential for a
non-linear dynamical system.  As has been shown quite clearly by the classical
field theory simulations of the early stages of reheating, the full non-linear
dynamics of even a purely classical system depends strongly on coherent effects
of backreaction due to the field's fluctuations\cite{classpreheat}. It is then
not unreasonable to expect that non-linear field fluctuations during inflation
could result in a departure from the dynamics derived directly from the
classical effective potential.

In this report we address these issues using the techniques of
out-of-equilibrium quantum field theory.  We examine a class of models in which
strongly quenched inflaton evolves under the influence of a negative mass
squared in the potential, a process which, following the terminology of such
phase transitions in condensed matter physics, we refer to as spinodal
decomposition\cite{condmatter}.  This class of inflation models includes
new\cite{new} and natural inflation\cite{natural}, as well as many models of
hybrid inflation\cite{hybrid}.  Such inflation models are of particular
interest in the present study because the evolution from the initial to the
final state of the system is necessarily a non-linear process.

The particular case in which the inflaton field is treated as a component of an
$O(N)$ vector in the large $N$ limit has already been detailed for the case of
new inflation, where it was found that the full non-perturbative quantum
dynamics does in fact reproduce an effectively classical trajectory for the
evolution of the inflaton.  The growth of quantum fluctuations results in a
dynamical flattening of the potential\cite{boyspin}, an analogue of the Maxwell
construction commonly used in studies of the equilibrium properties of phase
transitions.  In this earlier work, it was found that the effectively classical
trajectory of the inflaton field, a result of the phenomenon known as zero mode
reconstruction, is a consequence of Goldstone's theorem\cite{grav}.

However, the case of a single, real inflaton field is quite different.  
Here, the
long wavelength quantum fluctuations of the field also reassemble themselves
into a semi-classical field, but in this case the assembled
field {\it does not} obey the classical equations of motion of the original
potential.  Rather, the inflaton may be broken up into two components.  The
first is the mean field $\phi$ and obeys the classical equation of motion
expected from the original potential except that it is coupled to a second
field.  This second field $\sigma$, constructed through the assembly of quantum
fluctuations, obeys a modified equation of motion.  The result is an
effectively classical theory of two coupled fields, refered to as {\it spinodal
inflation}\cite{spinodal}.

The observational consequences are dramatic.  As there are effectively two
fields, the evolution becomes quite complicated and depends on the initial
conditions.  There are two regimes.  In the first, the mean inflaton field,
defined as the expectation value of the quantum field, has an initial value
greater than the expansion rate $\phi(t_0) > H_0/2\pi$ and the semi-classical
field $\sigma$ never becomes dynamically relevant.  The evolution reproduces
all of the standard results for that particular model of inflation and we can
think of this as the {\it classical} regime. In the second, quantum, regime for
which $\phi(t_0) < H_0/2\pi$, the influence of the $\sigma$ field is quite
important.  In this case, observational quantities such as the amplitude and
spectrum of primordial density fluctuations depend not only on the parameters
of the model, but also on the particular value of the initial mean field
$\phi(t_0)$.

A simple single field model can therefore produce a range of observational
results for any given choice of parameters.  In fact, due to the effective two
field dynamics, it is possible to produce observational features not possible
in the simple classical, single field version of the same theory, such as the
generation of a blue primordial power spectrum; this can occur much in the same
way as it does in hybrid inflation.

We begin with a short introduction to spinodal models of inflation and the need
for a fully out-of-equilibrium and non-perturbative description of the
dynamics.  We write down the self-consistent Hartree equations of motion for a
general spinodal potential, followed by an explanation of the assembly of
quantum fluctuations and how this results in an effectively classical two
field model.  Next, we move on to a detailed analysis of the two most important
spinodal models of inflation, $\lambda \Phi^4$ new inflation and natural
inflation.

Having determined the evolution of the field, we wish to examine the
observational consequences of spinodal inflation.  Beginning with the gauge
invariant formulation of gravitational perturbations of Mukhanov, Feldman, and
Brandenberger\cite{mfb}, we note that while the effective dynamics is that of
two fields, returning to the full quantum theory we see that only one set of
field fluctuations couples to gravity perturbations.

We then provide a complete and detailed numerical analysis of the dynamics of
spinodal inflation in single field models, including computations of the
primordial spectrum of scalar and tensor perturbations which result.  The
consequent $C_l$ spectra\cite{Cls} for a particularly interesting example 
is shown to allow for the direct relation of spinodal effects to observation.
Here we see explicitly the exciting features possible through the dynamics 
of spinodal inflation.


\section{Spinodal models of inflation}

\begin{figure}
\epsfig{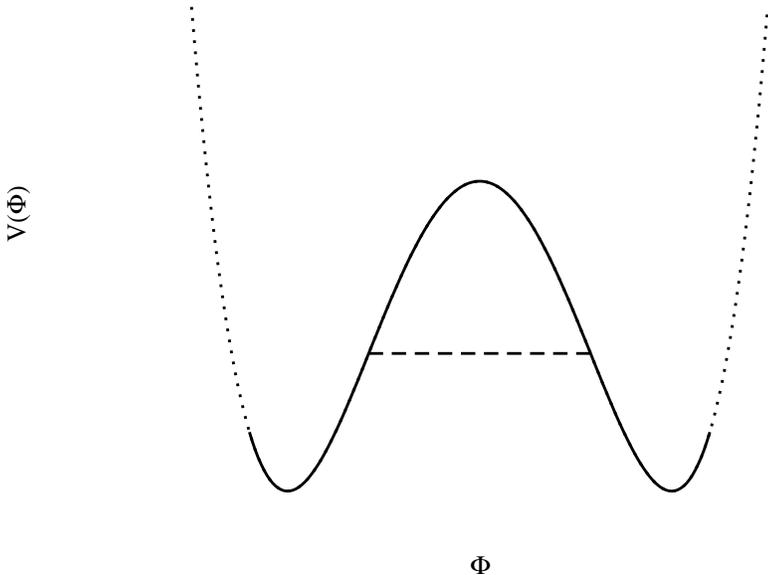}
\caption{A typical potential with a region of negative curvature.
The dashed line is the spinodal line, separating the unphysical
spinodal region (above the line) from the physical region (at and below
the line).  The exact shape of the dotted portion of the curves depends
upon the higher order terms in Eq.~(\ref{pot}).}
\label{spinpot}
\end{figure}

We envision the evolution of a scalar field near the top of a potential of the
form sketched in Fig.~\ref{spinpot}.  The potential may be expanded in the form
\begin{equation}
V(\Phi) = K - \frac12 \mu^2 \Phi^2 + \frac{\lambda}{4!} \Phi^4
+ \cdots \; , \label{pot}
\end{equation}
where the cosmological constant contribution $K$ is chosen such that the
potential is zero in the true vacuum, and $\mu^2$ is positive.  

Initially, the
field will ``roll'' slowly toward one of the minima of the potential.  This is
the regime in which inflation will take place.  To a first approximation, we
ignore the quartic term and see that the initial evolution follows that of a
free field in an inverted harmonic potential.  This evolution has been studied
in great detail in the context of inflationary cosmology\cite{guthpi}.  For
early times, the field grows exponentially. Eventually the
higher order terms in the potential become important, with the result that any
perturbative analysis of the dynamics will break down and must be augmented by
some non-perturbative technique.

Our choice of approximation is further restricted 
by the fact that the system we
wish to study is not in thermal equilibrium, thus leading us to real-time
methods. We emphasize that equilibrium constructs, such as the effective
potential, are completely inadequate tools for this problem.

The simplest approximation satisfying the requirements is the Gaussian
variational approximation, in which the quantum density matrix is restricted 
to take on a Gaussian form.  Also known as the Time Dependent Hartree-Fock
approximation, such mean field techniques have been utilized in quantum
mechanics dating back to Dirac\cite{dirac}.  It is a standard technique in
chemistry, condensed matter physics, and nuclear physics and has led to a
better understanding of the structure of a number of phase transitions.

\subsection{Real Scalar Field: Hartree Dynamics}

In what follows, we assume a spatially flat Robertson-Walker
metric:
\begin{equation}
ds^2 = dt^2 - a^2(t)d\vec{x}^2 \; .
\label{metric}
\end{equation}

We now derive the equations of motion for a real scalar field with Lagrangian
\begin{equation}
L = \frac12 \nabla^\mu \Phi(x) \nabla_\mu \Phi(x) -
V\left[\Phi(x);t\right] \; ,
\label{genlagrange}
\end{equation}
within the self-consistent Hartree approximation\cite{hartree}.  We break up
the field $\Phi$ into its expectation value plus a fluctuation about this
value:
\begin{eqnarray}
\Phi(\vec{x},t) &=& \phi(t) + \psi(\vec{x},t) \; , \\
\phi(t) & \equiv & \langle \Phi(\vec{x},t) \rangle \; .
\label{separate}
\end{eqnarray}
Here, $\phi$ depends only on time due to space translation invariance as is
consistent with the metric (\ref{metric}).  By definition $\langle
\psi(\vec{x},t) \rangle = 0$.

The Hartree approximation consists of replacing $\psi^{2n}$ by $c_1\langle
\psi^2 \rangle^{n-1}\psi^2 + c_2\langle \psi^2 \rangle^n$ and $\psi^{2n+1}$ by
$c_3 \langle \psi^2 \rangle^n \psi$, where the $c_i$ are constant factors whose
values are determined by Wick's Theorem.  This Hartree factorization may be
summarized as follows:
\begin{eqnarray}
\psi^{2n} &\to& \frac{(2n)!}{2^n (n-1)!} \langle \psi^2 \rangle^{n-1}\psi^2 
- \frac{(2n)!(n-1)}{2^n n!} \langle \psi^2 \rangle^n \; , \nonumber \\
\psi^{2n+1} &\to& \frac{(2n+1)!}{2^n n!} \langle \psi^2 \rangle^n \psi \; .
\label{factorization}
\end{eqnarray}

Given this factorization, any function $F(\Phi)$ becomes
\begin{equation}
F(\phi + \psi) = \sum_{n=0}^{\infty} \frac{1}{n!} \left(
\frac{\langle \psi^2 \rangle}{2} \right)^n \left\{
F^{(2n)}(\phi) + \psi F^{(2n+1)}(\phi) +\frac12 \left[ \psi^2 
- \langle \psi^2 \rangle \right] F^{(2n+2)}(\phi) \right\} ,
\label{factorF}
\end{equation}
where we use the notation
\begin{equation}
F^{(n)}(\phi) \equiv \frac{\delta^n}{\delta \phi^n} F(\phi) \; .
\end{equation}
Note that the latter two terms on the right hand side of
Eq.~(\ref{factorF}) have zero expectation value.  We therefore
find that the expectation value of a function factorizes as
\begin{equation}
\langle F(\phi + \psi) \rangle = \sum_{n=0}^{\infty} \frac{1}{n!} \left(
\frac{\langle \psi^2 \rangle}{2} \right)^n F^{(2n)}(\phi) \; .
\label{factorexpctF}
\end{equation}

The equations of motion for the mean field $\phi$ 
are given by the tadpole condition
$\langle \psi \rangle = 0$ \cite{tadpole}:
\begin{equation}
\ddot{\phi}(t) + 3\frac{\dot{a}(t)}{a(t)} + \sum_{n=0}^{\infty} 
\frac{1}{n!} \left(\frac{\langle \psi^2 \rangle}{2} \right)^n 
V^{(2n+1)}(\phi) = 0 \; ,
\label{phieqn}
\end{equation}
where we have used the metric (\ref{metric}) and have factorized the
expectation value $\langle V^{'}(\Phi) \rangle$ according to
Eq.~(\ref{factorexpctF}).  We define the Fourier transform of the Wightman
function by the expression
$$
G(\vec{x},t;\vec{x}^{'},t^{'}) = \int \frac{d^3k}{(2\pi)^3}
e^{i\vec{k}\cdot(\vec{x}-\vec{x}^{'})} G_k(t,t^{'}) \; ,
$$
where we have used the property of space translation invariance.
The quantity $\langle \psi^2(t) \rangle \equiv -i G(x,x)$ is constructed
from the mode functions obeying the equation
\begin{equation}
[G_k(t,t)]^{-1} f_k(t) = 0 \; ,
\end{equation}
together with the appropriate Closed Time Path boundary conditions\cite{ctp}.  
The operator $[G_k]^{-1}$
is given by the quadratic form appearing in the generating functional.
Explicitly, the $f_k(t)$ obey
\begin{equation}
\left[\frac{d^2}{dt^2} + 3\frac{\dot{a}(t)}{a(t)}\frac{d}{dt} + 
\frac{k^2}{a^2(t)} + \sum_{n=0}^{\infty} \frac{1}{n!} 
\left(\frac{\langle \psi^2 \rangle}{2} \right)^n V^{(2n+2)}(\phi) 
\right] f_k(t) = 0 \; ,
\label{fkeqn}
\end{equation}
where we have again used the factorization (\ref{factorexpctF}) to express
the potential term $\langle V^{''}(\Phi) \rangle$.

As mentioned above, the quantity $\langle \psi^2 \rangle$ is determined from
the mode functions $f_k$ combined with Closed Time Path boundary conditions
appropriate to the chosen initial state.  For an initial state in thermal
equilibrium at an initial temperature $T$, we have
\begin{equation}
\langle \psi^2(t) \rangle = \int \frac{d^3k}{(2\pi)^3} |f_k(t)|^2 
\coth \left(\frac{\omega_k}{2T}\right) \; ,
\label{psipsi}
\end{equation}
Note that in the zero temperature vacuum state given by the $T \to 0$ limit,
the hyperbolic cotangent has the value $1$.  The frequency $\omega_k$ appearing
here is given by
\begin{equation}
\omega_k^2 = k^2 + a^2(t_0) \sum_{n=0}^{\infty} \frac{1}{n!} 
\left(\frac{\langle \psi^2(t_0) \rangle}{2} \right)^n 
V^{(2n+2)}(\phi(t_0)) \; .
\label{omega_k}
\end{equation}
For the case of the $\lambda \Phi^4$ theory, there is an additional term 
proportional to the Ricci scalar $-a^2(t_0) R(t_0)/6$ which appears 
on the right hand side of this expression for $\omega_k$.
This term arises when one considers initial conditions corresponding to 
the adiabatic vacuum state in conformal time\cite{frwbig,init}, which 
is necessary if we wish our initial vacuum state to match the Minkowski 
vacuum in the limit $a(t) \to 1$.  However, this term is not necessary
if the scalar field is taken to be part of a low energy effective theory,
such as is the case with the natural inflation case we analyze below.

Using the vacuum state for the mode functions defined by the initial 
frequency spectrum of Eq. (\ref{omega_k}) leads to the following initial 
conditions on the $f_k(t)$:
\begin{eqnarray}
f_k(t_0) &=& \frac{1}{\sqrt{2 \omega_k}} \; , \nonumber \\
\dot{f}_k(t_0) &=& \left(-\frac{\dot{a}(t_0)}{a(t_0)} - i\omega_k \right) 
f_k(t_0) \; .
\label{initcond1}
\end{eqnarray}
A final note is that the initial frequencies $\omega_k$ given by 
Eq.~(\ref{omega_k}) may be imaginary for low $k$ modes.  In this 
case the initial conditions (\ref{initcond1}) need to be modified
for low $k$.  This may be done in a variety of ways with
little effect on results\cite{grav}.  Here, we choose a
smooth interpolation between low $k$ modes with modified frequencies and
the high $k$ modes which remain in the conformal vacuum state with
frequencies $\omega_k$ (\ref{omega_k}):
\begin{equation}
\omega_k^2 \equiv k^2+{\cal{M}}^2\tanh\left(
\frac{k^2+{\cal{M}}^2}{|{\cal{M}}^2|}\right) \; ,
\label{modinit}
\end{equation}
where 
\begin{equation}
{\cal M}^2 \equiv a^2(t_0) \sum_{n=0}^{\infty} \frac{1}{n!} 
\left(\frac{\langle \psi^2(t_0) \rangle}{2} \right)^n 
V^{(2n+2)}(\phi(t_0)) \; .
\end{equation}

This completes our set of equations of motion of the matter fields within the
Hartree approximation.

\subsection{Gravitational Dynamics}

We will treat gravity in the semi-classical approximation, in which the
expectation value of the full quantum energy-momentum tensor acts as a
classical source to the Einstein gravitational tensor.  The semi-classical
Einstein's equation reads
\begin{equation}
\frac{G^\mu_\nu}{8 \pi G_N}+\frac{\Lambda}{8 \pi G_N}+\left({\rm higher
derivatives}\right) = - \langle T^\mu_\nu \rangle \; ,
\label{semiclass}
\end{equation}
where $G_N = 1/M_{Pl}^2$ is (bare) Newton's constant, $\Lambda$ is the (bare)
cosmological constant and the components of the
Einstein curvature tensor using the metric (\ref{metric}) are:
\begin{eqnarray}
G^0_0 &=& -3 \frac{\dot{a}^2}{a^2} \; , \\
G^{\mu}_{\mu} &=& - R = -6 \left(\frac{\ddot{a}}{a} + 
\frac{\dot{a}^2}{a^2} \right) \; .
\end{eqnarray}
The higher derivative terms included in Eq.(\ref{semiclass}) are needed for
renormalization purposes; Newton's constant and the cosmological constant will
also be renormalized (see below). 

Once again, we use the factorization (\ref{factorexpctF}) to determine the
right hand side of (\ref{semiclass}).  Defining the additional integrals
\begin{eqnarray}
\langle \dot{\psi}^2(t) \rangle &=& \int \frac{d^3k}{(2\pi)^3} 
|\dot{f}_k(t)|^2 \coth \left(\frac{\omega_k}{2T}\right) \; , 
\label{dotpsipsi} \\
\langle (\vec{\nabla}\psi(t))^2 \rangle &=& \int \frac{d^3k}{(2\pi)^3} 
k^2 |f_k(t)|^2 \coth \left(\frac{\omega_k}{2T}\right) \; ,
\label{gradpsipsi}
\end{eqnarray}
we find for the energy 
density $\varepsilon$ and the trace ${\cal T}$ of the energy momentum 
tensor:
\begin{eqnarray}
\varepsilon &=& \langle T^0_0 \rangle = \frac12 \dot{\phi}^2 + 
\frac12 \langle \dot{\psi}^2 \rangle
+ \frac1{2a^2} \left\langle (\vec{\nabla}\psi)^2 \right\rangle
+ \sum_{n=0}^{\infty} \frac{1}{n!} \left(
\frac{\langle \psi^2 \rangle}{2} \right)^n V^{(2n)}(\phi) \; , 
\label{energy} \\ 
{\cal T} &=& \langle T^\mu_\mu \rangle = -\dot{\phi}^2 - \langle 
\dot{\psi}^2 \rangle + \frac1{a^2} \left\langle (\vec{\nabla}\psi)^2 
\right\rangle + 4 \sum_{n=0}^{\infty} \frac{1}{n!} \left(
\frac{\langle \psi^2 \rangle}{2} \right)^n V^{(2n)}(\phi) \; .
\label{trace}
\end{eqnarray}
The pressure $p$ is arrived at from these expressions through the relation $p =
(\varepsilon - {\cal T})/3$.  The equation of state of the system is
characterized by the quantity $w=p/\varepsilon$.

\subsection{Regularization and Renormalization}

The mode integrals appearing in Eq.~(\ref{psipsi}), (\ref{dotpsipsi}), and
(\ref{gradpsipsi}) are formally divergent and must be regularized in order to
perform any practical computation.

In the special case of a renormalizable potential $V(\Phi)$, for example in the
$\lambda \Phi^4$ model, we would like to fully renormalize the theory.  Our
choice of renormalization procedures is somewhat limited by the requirement
that the dynamics be amenable to numerical analysis. However, a number of
groups have recently addressed this problem either by means of adiabatic
regularization with a simple cutoff at large
momentum\cite{ramseyhu,frwbig,carmen}, as was first developed by Paul
Anderson\cite{anderson}, or by using a scheme based on dimensional
regularization\cite{baacke}.

In practice, we use the simple scheme developed by the Pittsburgh-Paris
collaboration.  This scheme has the advantage of being very easy to implement
and it has the attractive feature that all subtractions are absorbed into
counterterms renormalizing the bare couplings in the equation of motion and the
Friedmann equation (the latter must be extended to include a cosmological
constant and a higher order curvature term).  However, we mention that it does
not include the finite subtractions which would be necessary to give the
correct conformal anomaly.  These terms, the finite subtractions detailed in
\cite{ramseyhu} and \cite{carmen}, are formally important, but in the present
context, as the inflaton self-coupling is typically required to be of order
$10^{-12}$ or smaller, such terms will have absolutely no influence on
numerical simulations as their contributions are much smaller than the
numerical accuracy of the computations.  In fact, although we do not do so
here, it is normally safe to drop the logarithmically divergent terms from the
simulations as well without influencing the results.  The simulations are
checked after the fact to verify that they satisfy the covariant conservation
of the energy-momentum to within their numerical accuracy, and to ensure that
the results are independent of the value of the momentum cutoff.

For models with non-renormalizable potentials, such as natural inflation, we
have to satisfy ourselves with the treatment of the model as a low energy
effective theory with a well defined cutoff.  Again, we implement the
regularization by means of a large momentum cutoff.

\section{Early Time Dynamics and Reassembly}

As mentioned above, the initial linear dynamics in spinodal models of inflation
is well understood.  This period is characterized by exponential growth of both
the mean field $\phi$ and those mode functions $f_k$ with physical wavelength
greater than the Hubble distance, that is, $k/a(t) \ll H(t) \equiv
\dot{a}(t)/a(t)$.  Since $H(t)$ is approximately constant during inflation
while $a(t)$ is growing exponentially, clearly more and more modes satisfy this
condition at each subsequent time.  However, the very long wavelength modes
which crossed the horizon very early on in the evolution will tend to dominate
the quantity $\langle \psi^2 \rangle$ of Eq.~(\ref{psipsi}) simply because they
have experienced the spinodal instability for the longest time. This is a very
important characteristic of spinodal inflation which sets it apart from other
models: the dynamics is driven by a super-horizon scale quasi-particle
condensate.  On scales smaller than the horizon, it is not possible to
distinguish such a condensate from a purely homogeneous background field.  Any
possible measurement will determine only the combined properties of the
condensate and the mean field.  

Given the assumption of initial conditions near the local maximum of the
potential, and provided with a finite renormalized or regularized two-point
function along with very small values for the higher order couplings in the
Lagrangian, the early dynamics is well approximated by a linear analysis.  The
equation for the mean field for any potential of the form (\ref{pot}) is simply
\begin{equation}
\ddot{\phi}(t) + 3H(t)\dot{\phi}(t) - \mu^2 \phi(t) = 0 \; .
\label{linphi}
\end{equation}
To this order, we may take the Hubble parameter to be constant, $H(t)=H_0$, in
which case the solutions to this equation are simple exponentials.  Only the
growing term is relevant, so that we have the early time solution
\begin{equation}
\phi(t) \approx \phi_0 \exp\left[\left(\nu-\frac{3}{2}\right)H_0t\right] 
\quad \quad \nu = \sqrt{\frac{9}{4}+\frac{\mu^2}{H_0^2}} \; .
\label{linphisoln}
\end{equation}
The value of $\phi_0$ depends on the precise initial conditions for $\phi$ and
$\dot{\phi}$.

The mode functions obey the similar equation
\begin{equation}
\ddot{f}_k(t) + 3H(t)\dot{f}_k(t) + 
\left(\frac{k^2}{a^2(t)}-\mu^2\right) f_k(t) = 0 \; ,
\label{linfk}
\end{equation}
which for constant $H(t)=H_0$ and corresponding exponential $a(t) = \exp(H_0t)$
has the solutions
\begin{equation}
f_k(t) \approx \exp\left(-\frac{3}{2}H_0t\right) \left[
A_k J_{\nu}\left(\frac{k}{H_0}e^{-H_0t}\right) +
B_k J_{-\nu}\left(\frac{k}{H_0}e^{-H_0t}\right) \right] \;.
\label{linfksoln}
\end{equation}
The constants $A_k$ and $B_k$ are determined by the initial conditions on the
mode functions.  These solutions oscillate with an envelope proportional to
$1/a$ for sub-horizon modes (with $k/a > H_0$), but the solutions are growing
and decaying exponentials in the opposite limit $k/a \ll H_0$, leading to the
statement that super-horizon modes have an exponential instability.  In this
limit, we may again discard the exponentially decaying term.  We then have for
super-horizon modes
\begin{equation}
f_{k \ll aH}(t) \approx \frac{C_k}{k^{\nu-3/2}}
\exp\left[\left(\nu-\frac{3}{2}\right)H_0t\right] \; .
\label{linfksoln2} 
\end{equation}
Here, $C_k$ is roughly equal to the value of the mode $f_k$ evaluated at the
time $t(k)$ that the mode crosses the horizon as determined by the condition $k
= \exp[H_0 t(k)] H_0$.  For modes initially far within the horizon the general
dependence of $C_k$ on $k$ may be estimated from the initial behavior of the
mode functions $f_k \sim k^{-1/2}$ and the decay of the envelope of the Bessel
solutions which provides the standard result that $C_k \propto k^{-3/2}$.  We
note that while the exponential form (\ref{linfksoln2}) is only an asymptotic
solution for small arguments of the Bessel functions of Eq.~(\ref{linfksoln}),
due to the exponential behavior of this argument it very accurately describes
the evolution any mode function within a Hubble time of horizon crossing.

We are now in position to compute an expression for the condensate $\langle
\psi^2 \rangle$.  By separating the momentum space integral over super- and
sub-horizon modes respectively, we can take advantage of the expression
(\ref{linfksoln2}) to find for early times
\begin{equation}
\langle \psi^2 \rangle = \exp\left[\left(\nu-\frac{3}{2}\right)
H_0t\right] \int_0^{aH} \frac{d^3k}{2(2\pi)^3} 
\frac{|C_k|^2}{k^{2\nu-3}} + \langle \psi^2 \rangle_{k>aH} \; .
\label{psisep}
\end{equation}
The latter, sub-horizon term contains all subtractions due to renormalization.
After a few Hubble times, it is safe to neglect this term compared to the
exponentially growing super-horizon contribution.  To determine which modes are
most important to the evolution, it is convenient to examine the contribution
of each squared mode per logarithmic momentum interval, $dk/k$.  Using the
$k^{-3/2}$ behavior of $C_k$, we find for modes which are originally far inside
the horizon, but which have since crossed outside, that this contribution is
proportional to $k^{-(2\nu-3)}$.  Since $2\nu > 3$, the integral is dominated
by those modes which crossed the horizon the earliest.  This is important for
any numerical analysis as it allows one in practice to set a cutoff in the
calculation of $\langle \psi^2 \rangle$ with well controlled errors, avoiding
the problem of including a number of mode functions which grows exponentially
with the scale factor.

As a consequence of the formation of the condensate, it becomes possible to
form an accurate and simple model of the complete system, in which the full
two-point fluctuation $\langle \psi^2 \rangle$ is replaced by a nearly
homogeneous, and effectively classical field.  This produces a model in which
two effectively homogeneous classical fields, the mean field coupled to a
fluctuation field, accurately describe the dynamics.  The condensate field is
defined as
\begin{equation}
\sigma(t) \equiv \sqrt{\langle \psi^2(t) \rangle_{k<aH}} \; .
\label{sigmadef}
\end{equation}
For early times, it is given by the first term on the right hand side of
Eq.~(\ref{psisep}).  The square root of the value of the first integral of
(\ref{psisep}) a few Hubble times after the beginning of inflation provides an
effective initial condition on $\sigma$.  For a zero temperature initial state,
which for simplicity we take to be the case in what follows, this is found
numerically to be of order $H/2\pi$\cite{grav}.  The effect of a finite
temperature initial state is to increase this value by a factor of
$\sqrt{T/\mu}$ without modifying any of the qualitative features described in
this study.  The dynamics modelled by this classical field will be accurate up
to perturbatively small corrections to the dynamics due to the sub-horizon
modes contained in the final term of Eq.~(\ref{psisep}).

This system of effective homogeneous fields is referred to as a reassembled
system.  We will refer to the $\phi$ field, the expectation value of the full
quantum field, as the mean field, while we refer to the semi-classical $\sigma$
field as the fluctuation or condensate field.

While this condensate forms during the linear regime, eventually the dynamics
becomes non-linear.  It is this non-linear evolution which we wish to study.
We will be particularly interested in examining how the interaction of the
condensate with the mean field can influence the dynamics.

We consider primarily the dynamics of single field models (the large $N$ case
was studied previously\cite{grav}).  Here, the interactions of the condensate
and the mean field will be seen to lead to a complicated evolution in which
initial conditions play a primary role.

\section{Non-linear Dynamics}

We begin with an analysis of the stationary solutions for the mean field, which
obeys the equation of motion (\ref{phieqn}).  There are two primary late time
solutions.  The first is the trivial solution with $\phi(t)=0$, which will be
the solution for a system without symmetry breaking.  The other solution,
relevant for spinodal inflation, is given by the condition
\begin{equation}
\left\langle V'\left(\phi(t) + \psi(\vec{x},t)\right)\right\rangle_a
= \sum_{n=0}^{\infty} \frac{1}{n!} \left(
\frac{\langle \psi^2 \rangle_a}{2} \right)^n V^{(2n+1)}(\phi_a) = 0 \; ,
\label{sumruleH}
\end{equation}
where the subscript $a$ indicates the asymptotic solution.  Unlike the sum rule
in the large $N$ limit\cite{grav}, this condition does not correspond to
massless field modes\cite{destri}. 
Rather, for a bounded potential of the form (\ref{pot})
with a definite minimum at finite values for $\Phi$, the effective mass of
particle modes appearing in (\ref{fkeqn}) will be positive.  Asymptotically, we
therefore expect the field modes to be redshifted away due to expansion of the
universe such that the quantity $\langle \psi^2 \rangle$ becomes small and may
be neglected in (\ref{sumruleH}).  In this case, the expression for the
stationary solution becomes simply
\begin{equation}
V'(\phi_a) = 0 \; ,
\label{asymptH}
\end{equation}
and the effective mass of the field modes is
\begin{equation}
M^2_{eff} = V''(\phi_a) \; .
\label{asymptm2H}
\end{equation}
These, of course, are just the classical vacuum solutions in the symmetry
broken phase.

The task is to connect the early time solutions of Eqs.~(\ref{linphisoln}) and
(\ref{linfksoln}) and the late time solutions provided by Eqs.~(\ref{asymptH})
and (\ref{asymptm2H}).  It is convenient to introduce the effective condensate
field defined as in (\ref{sigmadef}).  Doing so, and neglecting the
exponentially suppressed gradient term in the expression for the energy density
(\ref{energy}), leads to the following equations of motion for the mean field
$\phi$, the condensate field $\sigma$, and the scale factor $a$:
\begin{eqnarray}
\ddot{\phi}+3\frac{\dot{a}}{a} \dot{\phi} + \sum_{n=0}^{\infty}
\frac{1}{2^n n!} \sigma^{2n} V^{(2n+1)}(\phi) & = & 0 \; , 
\label{assembly1} \\
\ddot{\sigma} + 3\frac{\dot{a}}{a} \dot{\sigma} + \sum_{n=0}^{\infty}
\frac{1}{2^n n!} \sigma^{2n+1} V^{(2n+2)}(\phi) & = & 0 \; , 
\label{assembly2}
\end{eqnarray}
\begin{equation}
\frac{\dot{a}^2}{a^2} = \frac{8\pi G_N}{3} 
\left[\frac12 \dot{\phi}^2 + \frac12 \dot{\sigma}^2
+ \sum_{n=0}^{\infty} \frac{\sigma^{2n}}{2^n n!} 
V^{(2n)}(\phi)\right] \; .
\label{assembly3}
\end{equation}
Remarkably, these equations are just those one would derive from a
classical system of two homogeneous fields with the potential
\begin{equation}
V(\phi,\sigma) = \sum_{n=0}^{\infty} \frac{1}{2^n n!} 
\sigma^{2n} V^{(2n)}(\phi) \; .
\label{2field}
\end{equation}

An important property of this potential is that, for interacting fields, it
does {\it not} depend symmetrically upon $\phi$ and $\sigma$.  
This means that, in
contrast to the large $N$ case\cite{grav}, it is not possible to combine the
mean and condensate fields into a {\it single} effective classical inflaton.  
The fields $\phi$ and $\sigma$ have different dynamics and properties.

The initial condition $\sigma(t_0) \simeq H_0/2\pi$ results in two distinct
regimes.  In the classical regime, characterized by $\phi(t_0) \gg H_0/2\pi$
the dynamics is dominated by the evolution of $\phi$ and is effectively
independent of $\sigma$. This follows from the fact that $\sigma$ never grows
to be particularly large before $\phi$ reaches its classical minimum.  However,
in the fluctuation dominated regime where $\phi(t_0) \ll H_0/2\pi$, $\sigma$
has a significant effect on the evolution of $\phi$ and dramatically modifies
the overall inflationary dynamics from naive expectations.

Before moving on to specific examples, 
let us examine some general features of the
fluctuation dominated regime.  At intermediate times the dynamics of the
reassembled fields will be primarily dictated by the equation of motion of
$\sigma$, due to the fact that $\sigma \gg \phi$ 
and that both fields contribute
to the equations of motion in a similar way.  The dynamics will therefore
approach a quasi-equilibrium regime for which the third term in
Eq.~(\ref{assembly2}) becomes small:
\begin{equation}
V_{,\sigma}(\phi,\sigma) \to 0 \; ,
\label{sumrule}
\end{equation}
where the comma represents the partial derivative.  Returning to the equation
of motion for the mode functions (\ref{fkeqn}), we discover that this condition
corresponds to effectively massless quanta.  What we are seeing is the rapid
departure of the inflaton field from the unstable regime for which its field
quanta have a negative mass squared into a physical regime of massless quanta.
This flattening of the unphysical regime into a form for which particles become
well defined is reminiscent of the famous Maxwell Construction describing the
convexity of the thermodynamical equilibrium free energy of a system.  The
behavior described here is the out of equilibrium analog of such a
construction\cite{boyspin}.

We note, however, that this condition (\ref{sumrule}) does not correspond to
the late time classical solutions (\ref{sumruleH}) we expect; it is 
only a quasi-equilibrium for which the mean field $\phi$ continues to evolve
toward the true minimum, as is clear from Eq.~(\ref{assembly1}).

To understand the implications of this behavior better, it is useful to work
with definite models.  We now turn to concrete examples which will allow us to
follow the dynamics in detail.

\subsection{New Inflation} \label{new}
The simplest spinodal model of inflation is new inflation for which the scalar
potential (\ref{pot}) is truncated at quartic order with the cosmological
constant contribution $K = 3\mu^4/2\lambda$:
\begin{equation}
V(\Phi) = \frac{3\mu^4}{2\lambda} - \frac12 \mu^2 \Phi^2 
+ \frac{\lambda}{24}\Phi^4 \; .
\label{phi4pot}
\end{equation}
This model is of particular importance due to its renormalizability.

\begin{figure}
\epsfig{file=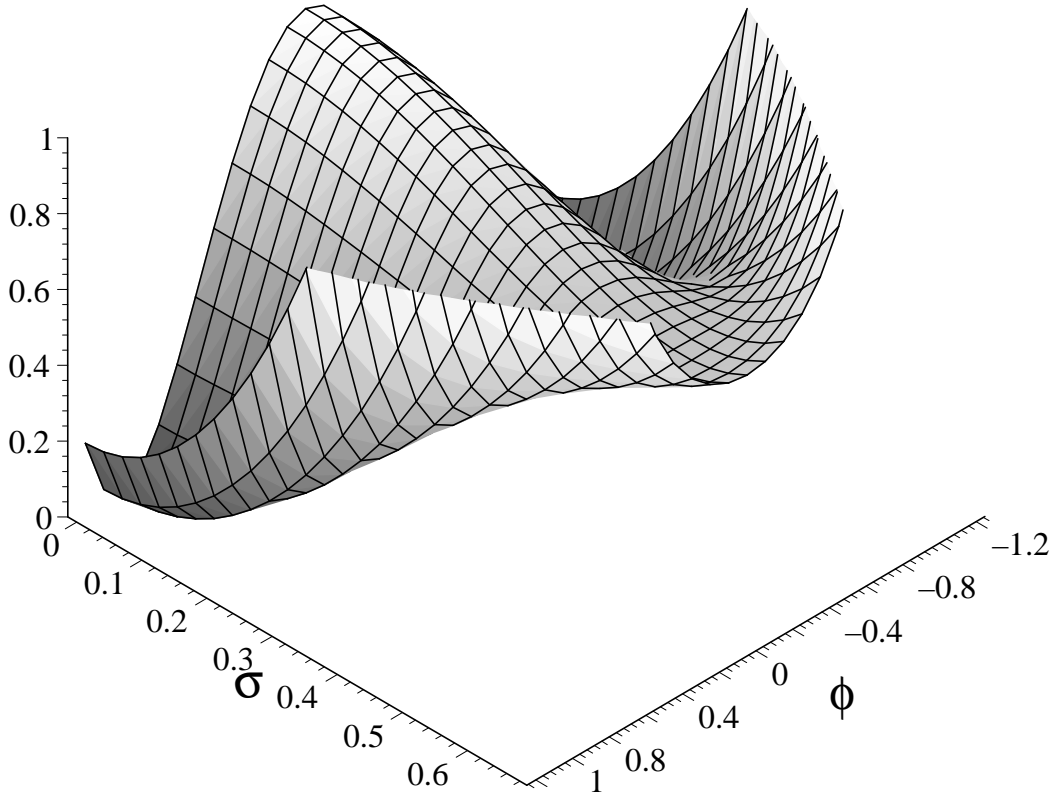}
\caption{The two dimensional potential which may be used to describe
the full non-linear field dynamics in a $\lambda \Phi^4$ new inflation 
phase transition in terms of effectively classical fields $\phi$ and 
$\sigma$.  Axes are scaled such that the true minima of the potential 
occur at $\phi=\pm 1$.}
\label{phipot}
\end{figure}

The reassembled equations of motion are
\begin{eqnarray}
\ddot{\phi}+3\frac{\dot{a}}{a} \dot{\phi} - \mu^2 \phi 
+ \frac{\lambda}{6}\phi^3 + \frac{\lambda}{2}\sigma^2 \phi & = & 0 \; , 
\label{lphiassembly1} \\
\ddot{\sigma} + 3\frac{\dot{a}}{a} \dot{\sigma} - \mu^2 \sigma
+ \frac{\lambda}{2}\sigma^3 + \frac{\lambda}{2}\phi^2 \sigma & = & 0 \; , 
\label{lphiassembly2}
\end{eqnarray}
\begin{equation}
\frac{\dot{a}^2}{a^2} = \frac{8\pi G_N}{3} 
\left[\frac12 \dot{\phi}^2 + \frac12 \dot{\sigma}^2 
+ \frac{3 \mu^4}{2\lambda} - \frac12 \mu^2 \left(\phi^2+\sigma^2\right)
+ \frac{\lambda}{24}\left(\phi^4+3\sigma^4+6\phi^2 \sigma^2\right) 
\right] \; .
\label{lphiassembly3}
\end{equation}
The reassembled two field potential for this model is therefore
\begin{equation}
V(\phi,\sigma) = \frac{3 \mu^4}{2\lambda} - \frac12 \mu^2 
\left(\phi^2+\sigma^2\right)
+ \frac{\lambda}{24}\left(\phi^4+3\sigma^4+6\phi^2 \sigma^2\right) 
\; .
\label{lphiassembly4}
\end{equation}
A plot of this two dimensional potential is shown in Fig.~\ref{phipot}.  
It is characterized by a local maximum at $\phi=0$, $\sigma=0$, a saddle 
point at $\phi=0$, $\sigma = \mu \sqrt{2/\lambda}$, and minima at 
$\phi=\pm \mu \sqrt{6/\lambda}$, $\sigma = 0$.

As we discussed above, there are two dynamical regimes determined by
the initial value of $\phi$.  In the classical regime, $\sigma$ never
plays a significant role and the field $\phi$ acts as an ordinary 
classical field in a $\lambda \phi^4$ potential (\ref{phi4pot}).  
However, in the fluctuation dominated
regime, the evolution proceeds first through an inflationary phase 
with energy density contribution given by the tree level potential,
but then enters a second regime for which the condition (\ref{sumrule})
is satisfied.
 
For this potential, we have in the quasi-equilibrium regime the sum rule
\begin{equation}
-\mu^2 + \frac{\lambda}{2}\sigma^2 + \frac{\lambda}{2}\phi^2 = 0 \; ,
\label{sumrulelphi4}
\end{equation}
which we recognize as the condition for massless quanta.
This expression may then be substituted
back into the equation of motion for the $\phi$ field 
(\ref{assembly1}), where we find 
\begin{equation}
\ddot{\phi}+3\frac{\dot{a}}{a} \dot{\phi} - \frac{\lambda}{3}\phi^3 = 0 \; .
\label{phisumruled}
\end{equation}
Here we see that the potential energy contribution to the equation 
of motion for $\phi$ appears at the cubic order.  The field $\phi$ 
therefore evolves as a field with an effective squared mass  
given by $-\lambda\phi^2/3$.  As this is typically 
much smaller (in absolute value) than $\mu^2$, what we observe is 
that the potential becomes flattened as a consequence of the 
non-perturbative growth of fluctuations.  

We can also use the condition (\ref{sumrulelphi4}) to determine 
the value of the potential energy (\ref{lphiassembly4}) during this 
phase.  We find simply
\begin{equation}
V(\phi,\sigma) \to \frac{\mu^4}{\lambda} - \frac{\lambda}{12}\phi^4
\; ,
\label{spinodalenergy}
\end{equation}
as is consistent with the effective mass for $\phi$.  We see that
the growth of fluctuations has produced an effective dynamics 
corresponding to a very flat potential for $\phi$ with a cosmological
constant contribution to the energy density $\mu^4/\lambda$ which
is $2/3$ of the value appearing in the original potential.  We 
therefore arrive at a second stage of inflation with an expansion
rate related to the original stage by a factor of $\sqrt{2/3}$.

We note, however, that there is a continued instability in $\phi$ with
the consequence that eventually the condition 
\begin{equation}
\phi^2 \geq \phi_s^2 = 2\mu^2/\lambda \; ,
\label{spinodalpoint}
\end{equation}
is met and it becomes impossible to 
satisfy the sum rule (\ref{sumrulelphi4}).  At this point, the
second inflationary stage ends, the fluctuation field $\sigma$ 
decays away and the field $\phi$ evolves to its classical minimum.


\subsection{Natural Inflation} \label{natural}
The natural inflation potential is derived from the vacuum manifold
of a complex scalar field spontaneously broken at the Planck scale
and with explicit symmetry breaking at the Grand Unified scale. 
It may be written in the form
\begin{equation}
V(\Phi) = \Lambda^4 \left[1+\cos(\Phi/f)\right] \; ,
\end{equation}
where $\Lambda \sim M_{GUT}$ and $f \sim M_{Pl}$ 
are constants.  Expansion of the cosine reveals that this potential is
of the form (\ref{pot}) with $\mu^2 \equiv \Lambda^2/f$ and 
$\lambda \equiv \Lambda^4/f^4$.

The reassembled equations of motion become:
\begin{eqnarray}
\ddot{\phi} + 3H \dot{\phi} - \frac{\Lambda^4}{f} 
\sin\left(\frac{\phi}{f}\right) 
\exp\left(-\frac{\sigma^2}{2f^2}\right) &=& 0 \; , 
\label{natassembly1} \\
\ddot{\sigma} + 3H \dot{\sigma} - \frac{\Lambda^4}{f^2} 
\cos\left(\frac{\phi}{f}\right) 
\exp\left(-\frac{\sigma^2}{2f^2}\right) \sigma &=& 0 \; ,
\end{eqnarray}
\begin{equation}
\frac{\dot{a}^2}{a^2} = \frac{8\pi G_N}{3} \left[ \frac12 \dot{\phi}^2 
+ \frac12 \dot{\sigma}^2 + \Lambda^4 \left[1+\cos\left(\frac{\phi}{f}\right) 
\exp\left(-\frac{\phi^2}{2f^2}\right)\right] \right]
\; . 
\end{equation}

\begin{figure}
\epsfig{file=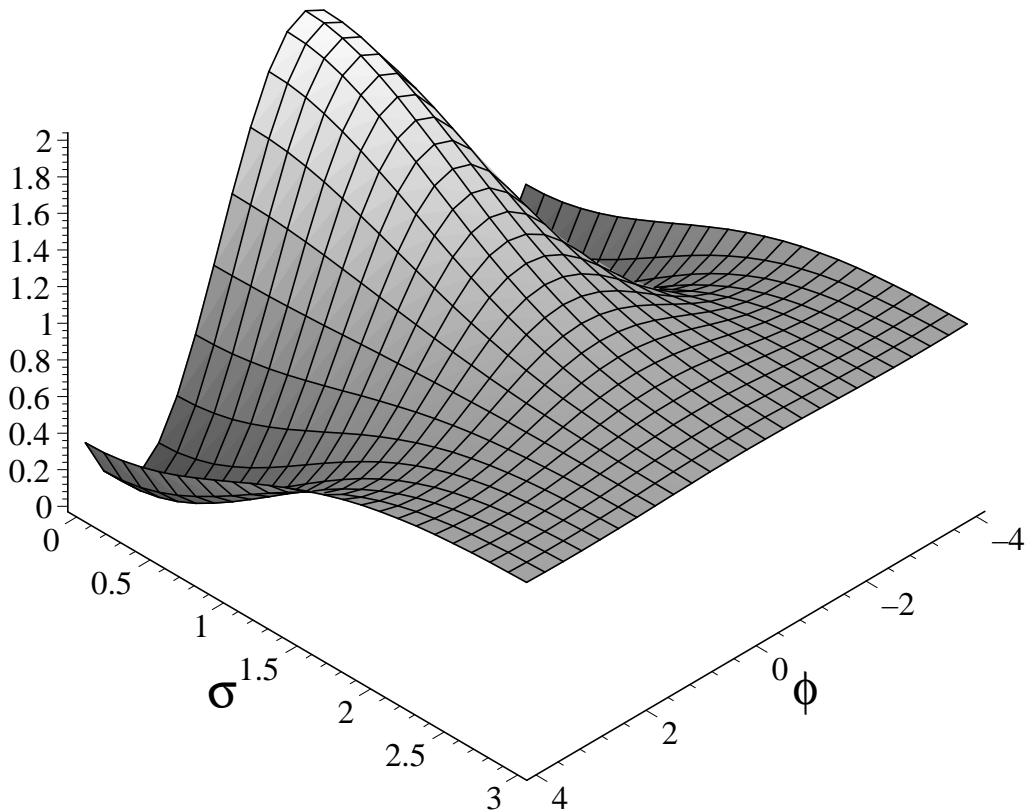}
\caption{The two field classical potential for natural inflation.  
Axes are scaled such that the true minima of the potential 
occur at $\phi = (2j+1) \pi$.}
\label{natpot}
\end{figure}

We recognize these equations as those of a two homogeneous
classical scalar fields with potential 
\begin{equation}
V(\phi,\sigma) = 
\Lambda^4 \left[1+\cos\left(\frac{\phi}{f}\right) 
\exp\left(-\frac{\sigma^2}{2f^2}\right)\right]
\; . 
\label{Heffpot}
\end{equation}
We provide a plot of this two field potential in Fig.~\ref{natpot}.
As expected, for any integer $j$, we have degenerate maxima at 
$\phi = 2j \pi f$, $\sigma = 0$ and degenerate minima at 
$\phi = (2j+1) \pi f$, $\sigma =0$.  The feature to notice, however,
is that the potential quickly becomes very flat as $\sigma/f$ becomes
greater than $1$.  

Again, we concern ourselves with the fluctuation dominated regime.
We find that the sum rule (\ref{sumrule}) results in the condition
\begin{equation}
\frac{\Lambda^4}{f^2} \cos\left(\frac{\phi}{f}\right) 
\exp\left(-\frac{\sigma^2}{2f^2}\right) \to 0 \; ,
\label{natsumrule}
\end{equation}
which for $\phi < \pi f $ becomes satisfied as $\sigma$ grows 
large.  The effective mass term for $\phi$ in (\ref{natassembly1})
becomes exponentially suppressed as well, again indicating that
the growth of fluctuations of the field results in a flattening
of the potential.  The potential itself clearly goes to the
value $\Lambda^4$ which is precisely half of the value of the
original cosmological constant contribution.  This state of 
affairs will continue until $\phi \geq \phi_s = \pi f$, at which point 
the fluctuations represented by $\sigma$ become massive and 
begin to decay away.

As in the $\lambda \Phi^4$ model, we expect two stages of inflation,
this time with the expansion rate of the second stage reduced from that
of the first by the factor $\sqrt{2}$.  

\section{Metric Perturbations}

The possible link between inflationary expansion of the universe at Grand
Unified energy scales and the observation of fluctuations in the Cosmic
Microwave Background (CMB) temperature of order $10^{-5}$ is 
remarkable\cite{COBE}.  The
fact that we are now in the process of observing the details of this
temperature spectrum through a number of ground-, air-, and space-based
experiments, presents an amazing opportunity for probing details of what the
universe was like at times inaccessible through any other means\cite{CMBrev}.

In order to take full advantage of this opportunity, however, it is important
that we are careful in connecting the observations to the theoretical models of
the dynamical processes of the Early Universe which may have led to the CMB
anisotropies.  We have already presented a detailed 
analysis of the cosmological
dynamics of scalar fields undergoing inflationary phase transitions. Here, we
provide computations of the primordial spectra of density perturbations which
arise from such phase transitions.

By computing the spectrum of scalar perturbations, we explicitly show that
coherent effects due to large wavelength fluctuations of the inflaton field
could have significant impact on observational features of the temperature
spectrum of CMB anisotropies.  Particular features of interest include scales
on which there is deviation from a flat Harrison-Zel'dovich primordial power
spectrum with power increasing as one moves to smaller scales.  Such a
spectrum, referred to as having a blue tilt (corresponding to a scalar tilt
parameter $n_s > 1$), was previously thought only to be possible in more
complicated, multi-field models of inflation.  Here, we find that it is indeed
possible to produce such a spectrum on the length scales relevant to CMB
anisotropy observations from the simplest of single field models.

Another significant feature of the spectra produced in phase transitions is the
possible decrease in the amplitude of the primordial perturbations on the
scales of interest relative to the predictions of an analysis assuming a
classical evolution for the inflationary field.  This may somewhat alleviate
the fine tuning problem, although we find in practice that this effect is
relatively minor, allowing perhaps a dimensionless quartic coupling for the
field one or two orders of magnitude larger than previously thought.  Since
this coupling is typically thought to be restricted to be less than $10^{-14}$,
we still require the inflationary models to be highly fine tuned.

A final feature of the spectra is their dependence on the precise initial state
of the inflationary field in the region corresponding to today's observational
universe.  It is found that a universe which began the inflationary phase in a
strongly classical state will show none of the exotic features described here.
This strongly contrasts with the case of initial states in which quantum
fluctuations of the field are of the same order as the classical field value.
In this latter case, the primordial spectrum may depend quite strongly on
precisely how `quantum' the particular initial state is.

We begin with a computation of the amplitude of scalar and tensor metric
perturbations resulting from spinodal inflation.  For the specific models
discussed above, we provide details of the resulting power spectrum as a
function of scale.  We also include plots of the amplitude and tilt in the
power spectrum as a function of the initial state of the scalar field for a
choice of scale consistent with those observed by the Cosmic Background
Explorer\cite{COBE}.  Finally, we provide examples of the $C_l$ temperature 
anisotropy spectra\cite{Cls} that result from particularly 
interesting examples of spinodal
inflationary effects to allow comparison with other theoretical plots of the
$C_l$ spectrum and with the observed spectrum.

Note that in what follows, we use the normalizations of the scalar and tensor
amplitudes of Ref.~\cite{mfb}.

\section{The Primordial Spectrum}
Our starting point for the computation of primordial density perturbations is
the expression for the average energy density (\ref{energy}).  To compute the
average fluctuation we define
$$
\delta \varepsilon \equiv \left\langle \left(\delta T_0^0\right)^2
\right\rangle^{1/2} \; ,
$$
where the variation of the energy density yields the expression
\begin{equation}
\left(\delta T_0^0\right) = \left(\dot{\phi} + \dot{\psi}\right)
\dot{\psi} + \frac{1}{a^2} \left(\vec{\nabla} \psi \vec{\nabla}\right)
\psi + \frac{\delta V(\phi+\psi)}{\delta \psi} \psi \; .
\end{equation}
It is convenient to introduce the Fourier variable
$$
\psi_k = \frac{k^{3/2}}{\sqrt{2}\pi} f_k \; ,
$$
defined such that
$$
\langle \psi^2 \rangle = \int \frac{dk}{k} |\psi_k|^2 \; .
$$
In Fourier space, we find the following expression for the average energy
density fluctuation on a scale corresponding to a wavenumber $k$:
\begin{eqnarray}
\left\langle \left(\delta T_0^0\right) \left(\delta
T_0^0\right) \right\rangle_k &=& |\dot{\psi}_k|^2 \left(
\dot{\phi}^2 + \langle \dot{\psi}^2 \rangle \right) + \left(
\psi_k \dot{\psi}_k^* + \dot{\psi}_k \psi_k^* \right) \left\langle
\left( (\dot{\phi} + \dot{\psi} \right) \frac{\delta
V(\phi+\psi)}{\delta \psi} \right\rangle \nonumber \\
&+& |\psi_k|^2 \left\langle \left( \frac{\delta V(\phi+\psi)}{\delta
\psi} \right)^2 \right\rangle \; ,
\label{deltaT00sqr}
\end{eqnarray}
where terms proportional to $1/a^2$ have been neglected. Normally during
inflation, the third term on the right hand side of (\ref{deltaT00sqr})
dominates, which is equivalent to the statement that the usual slow roll
conditions are satisfied during the evolution.

Using the above expression, the density contrast at first horizon 
crossing is
\begin{equation}
|\delta_{k=aH}|^2 = \frac{1}{\rho^2} \left\langle \left(\delta
T_0^0\right) \left(\delta T_0^0\right) \right\rangle \; .
\label{delta1st}
\end{equation}
Here, we note that we have used the important relation between the
gauge invariant Bardeen variable representing the scalar metric perturbation 
and the energy density fluctuation at horizon crossing to allow us to write
down this apparently simple expression.
As the density perturbations are adiabatic (recall that there is
only one scalar field with one set of fluctuations), the super-horizon 
evolution of the metric perturbation up to the second horizon crossing 
is specified by the conservation of a single parameter $\xi$ as defined
in Ref.~\cite{mfb}.
Following the procedure of Ref.~\cite{mfb}, we arrive at the
expression for the density contrast at mode re-entry in terms of the scaled
mode functions $\psi_k$:
\begin{equation}
|\delta_h(k)|^2 = \frac{\left[\left(\dot{\phi}^2 + \langle
\dot{\psi}^2 \rangle\right) |\dot{\psi}_k|^2 + \left\langle
(\dot{\phi} + \dot{\psi}) \frac{\delta V(\phi+\psi)}{\delta \psi}
\right\rangle (\psi_k \dot{\psi}_k^* + \dot{\psi}_k \psi_k^*) + \left\langle
\left( \frac{\delta V(\phi+\psi)}{\delta \psi} \right)^2
\right\rangle |\psi_k|^2 \right]}{25 \left( \dot{\phi}^2 +
\langle \dot{\psi}^2 \rangle\right)^2} \; ,
\label{deltaH}
\end{equation}
where each quantity is evaluated at the time when the
corresponding mode first crosses the horizon, i.e. when $k=aH$.

In what follows, we will assume that the third term in
(\ref{deltaT00sqr}) dominates over the first two terms, as this is
an excellent approximation in the models in which we are
interested.  Eq.~(\ref{deltaH}) becomes simply:
\begin{equation}
|\delta_h(k)|^2 \simeq \frac{|\psi_k|^2 \left\langle \left(
\frac{\delta V(\phi+\psi)}{\delta \psi} \right)^2
\right\rangle}{25 \left(\dot{\phi}^2 + \langle \dot{\psi}^2
\rangle\right)^2} \; .
\label{deltaH2}
\end{equation}

This can be simplified further by recognizing that at first
horizon crossing the quantity $|\psi_k|^2$ is given approximately
by $(H/2\pi)^2$\cite{H2pi}.  This may be seen directly from the asymptotic
solutions of the mode functions for large momenta.  As the
expansion is rapid, these asymptotic solutions remain
approximately valid all the way to $k \sim aH$.  Combining this
with the (semi-classical) Friedmann equation and using the inflationary
condition $\langle V \rangle \gg \dot{\phi}^2 + \langle \dot{\psi}^2 \rangle$,
we reach the result
\begin{equation}
|\delta_h(k)|^2 \simeq \frac{2}{75\pi} \frac{\langle V \rangle \langle
V_{,\psi}^2 \rangle}{M_{Pl}^2 (\dot{\phi}^2 + \langle \dot{\psi}^2
\rangle )^2} \; .
\label{deltaH3}
\end{equation}
If we were to make the further assumption that the fluctuations
are always small with $\psi \ll \phi$ and $\dot{\psi} \ll
\dot{\phi}$, then we would arrive at the standard slow roll
expression for the density contrast:
\begin{equation}
|\delta_h(k)| =  \sqrt{6\pi} \frac{8}{5} 
\frac{V^{3/2}}{|V'| M_{Pl}^3} \; .
\label{slowrolldeltaH}
\end{equation}
We will continue, however, with the more general form
(\ref{deltaH3}).

The computation of the tilt parameter $n_s-1$ is straightforward,
given (\ref{deltaH3}).  We define
\begin{equation}
n_s-1 \equiv \left. \frac{d (\ln|\delta_h(k)|)}{d \ln(k)}\right|_{k=aH} \; .
\label{nsminusone}
\end{equation}
As is common practice, we could rewrite this expression in terms
of partial derivatives of the field variables.  However, since we
have dependence both upon $\phi$ and the fluctuations $\psi$, such
a procedure results in a complicated expression which is not
particularly instructive.  We will therefore satisfy ourselves
with (\ref{nsminusone}), which is used directly to compute the
tilt parameter in the numerical examples.

\subsection{$\lambda \Phi^4$ Spinodal Inflation}
Using (\ref{deltaH3}), we now compute the expression for the primordial
spectrum of scalar metric perturbations specific to the $\lambda \Phi^4$ model
of spinodal inflation.  In terms of the reassembled variables $\phi$ and
$\sigma$, the result is:
\begin{eqnarray}
|\delta_h(k)| &=& \sqrt{\frac{2}{75\pi}} \frac{1}{M_{Pl} \left(\dot{\phi}^2 + 
\dot{\sigma}^2\right)} \left[ \frac{3 \mu^4}{2\lambda} 
- \frac12 \mu^2 \left(\phi^2+\sigma^2\right)
+ \frac{\lambda}{24}\left(\phi^4+6\phi^2 \sigma^2+3\sigma^4\right)
\right]^{1/2} \nonumber \\
& \times & \left[\mu^4 \left(\phi^2+\sigma^2\right) 
- \frac{\lambda \mu^2}{3}\left(\phi^4 + 6\phi^2\sigma^2 + 3 \sigma^4
\right)\right. \nonumber \\
& & \left. + \frac{\lambda^2}{36} \left(\phi^6 + 15\phi^4\sigma^2
+ 45\phi^2\sigma^4 + 15\sigma^6\right)\right]^{1/2}
\; .
\label{lphi4deltaH}
\end{eqnarray}
Again, we point out that all expressions are to be evaluated when the
given scale $k$ crosses the horizon.

\subsection{Natural Inflation}
For natural inflation, the relevant expression is:
\begin{equation}
|\delta_h(k)| = \frac{\Lambda^6}{\sqrt{75\pi}M_{Pl}f\left(\dot{\phi}^2 + 
\dot{\sigma}^2\right)} 
\left( 1 - \cos\left( 2\phi/f \right)
e^{-2\sigma^2/f^2}\right)^{1/2} \left(1 + \cos\left(\phi/f\right)
e^{-\sigma^2/2f^2} \right)^{1/2} \; .
\label{natdeltaH}
\end{equation}

\subsection{Gravitational Wave Perturbations}
It is also of interest to examine the spectrum of gravitational wave
perturbations resulting from spinodal inflation.  As such perturbations do not
directly interact with the inflaton field, they may be related directly to the
expansion rate during any inflationary stages.  During such regimes, the
amplitude of gravitational waves is simply\cite{mfb}:
\begin{equation}
|\delta_g(k)| = \frac{2}{\sqrt{3\pi}} \frac{H_k}{M_{Pl}} \; ,
\label{deltaG}
\end{equation}
where $H_k$ is the value of the expansion rate when the scale $k$ first crosses
the horizon, $k = a H_k$.

As we have discussed, spinodal inflation may involve two distinct inflationary
stages.  The relevant amplitude of the gravitational waves is therefore
typically determined by which stage is in effect 60--50 $e$-folds before the
end of inflation, when the relevant length scales exit the horizon.  However,
the transition period between inflationary stages may also be relevant.

The spectrum in the transition period depends upon the details of the
transition, but clearly must smoothly interpolate between the two major
regimes. The important factor is whether the characteristic accelerated
expansion continues to hold throughout the transition or if the transition
includes a short period of deceleration.  In the former case, the transition is
smooth and straightforward, following Eq.(\ref{deltaG}) throughout, while in
the latter case, the amplitude of the modes crossing the horizon during the
transition has an oscillatory nature\cite{mfb}.

As we will see, the former, smooth transition without any oscillation of the
gravitational wave amplitude is typical of spinodal inflation, with the result
that the spectrum may have at most a single feature indicative of the
transition from the initial inflationary phase to the spinodal phase.

\subsection{Notes on metric perturbations}
It is worth taking a moment to examine the significance of the
expressions for the amplitudes of the metric perturbations in some
detail.  The first feature to notice is that the scalar amplitude
depends directly upon both the average field value $\phi$ and the
typical fluctuation represented by $\sigma$.  As we have seen, the
end of inflation depends upon the evolution of $\phi$, occurring
when $\phi$ reaches the spinodal value $\phi_s$.  When the
influence of the field fluctuations are neglected, the value of
$\phi$ $60$ $e$-folds before the end of inflation, $\phi_{60}$, which
we will take to be the largest scale measured by COBE, is
a well defined quantity and corresponds to a well defined scalar
amplitude $|\delta_{60}|$ independent of the initial conditions.

However, we see here that the dynamics of the field fluctuations
$\sigma$ may influence the evolution of $\phi$.  The net result is
that the precise values of $\phi_{60}$, $\sigma_{60}$, and
therefore $|\delta_{60}|$ occuring when the relevant mode crosses
the horizon does generally depend upon the initial conditions for
the inflaton.

This leads to an extremely important result:  {\it The spectrum of
primordial metric perturbations resulting from a given model of
inflation depends upon the initial state of the inflaton field.}

A second significant feature coming as a result of these more
general expressions for the perturbation amplitude is that
$|\delta_h(k)|$ is not necessarily a monotonically increasing
function of length scale.  In contrast to the case in which the
influence of the field fluctuations are neglected, there may be
periods of time during which $|\delta_h(k)|$ increases as ever
shorter length scales cross the horizon.

This leads to a result previously thought to be impossible in such
simple, single field inflationary models: {\it Inflation may
result in a blue spectral tilt in the primordial spectrum of
scalar perturbations.}

\subsection{The $C_l$ Spectrum}
As a final note in this section, we recall that none of the parameters of the
primordial spectrum written down to this point are directly observable.
Rather, this primordial spectrum becomes processed by the evolution of the
late time universe, after corresponding length scales have re-entered the
particle horizon.  The most important set of observable parameters is the
$C_l$ spectrum of the Cosmic Microwave Background map of temperature
variations.

Standard techniques of computing the $C_l$'s from the primordial parameters
assume a scale independent tilt $n_s$, which does not generally apply here.
It is, of course, possible to compute the $C_l$ spectrum for a more general
primordial spectrum, but to do so is far beyond the scope of the present
work.  We therefore use a standard approximation which relates the tilted 
spectrum $C_l^{(tilt)}$ to the tilted scalar perturbation amplitude 
$|\delta_h(k)^{tilt}|$ evaluated on the scale $k=l H_*/2$ where $H_*$ is the
inverse Hubble radius today relative to the respective quantities where a flat
spectrum ($n_s=1$) is assumed\cite{LiddleLyth}:
\begin{equation}
C_l^{(tilt)} = C_l^{(flat)} 
\frac{|\delta_h(k=l H_*/2)^{tilt}|^2}{|\delta_h(k=l H_*/2)^{flat}|^2} \; .
\label{tiltedCls}
\end{equation}
Assuming the spectrum $C_l^{(flat)}$ used here is properly COBE normalized, we
require that the flat and tilted primordial amplitude match for the low $l$ 
scales corresponding to COBE.

This will allow us to present an approximate plot of the $C_l$ spectrum
resulting from spinodal effects which may be compared to standard $C_l$
spectra.

\section{Concrete examples and results}
We now put all the pieces of the previous sections together in numerical
simulations of the full dynamical equations of motion of the spinodal system.
A note is in order regarding these simulations. In certain cases it was
impractical to run numerical simulations using the full field theory equations
of motion.  These include plots showing quantities as a function of the initial
condition for $\phi$ as well as the plot of the scalar tilt as a function of
the initial Hubble parameter.  In these cases, we turned to the classical
2-field models described above, using the approximate initial condition
$\sigma(t_0) = H_0/2\pi$.  All such figures specify in the caption that they
result from the classical models.  All other figures were produced from full
field theory simulations.

\subsection{New Inflation}
We begin with the $\lambda \Phi^4$ system, which we first examined in this
context in Ref.~\cite{spinodal}.  We show the dynamics of the mean field
$\phi$, the fluctuation $\langle \psi^2\rangle^{1/2}$, each of which is scaled
by the factor $f \equiv \mu \sqrt{6/\lambda}$, and the expansion rate $H$ for
each of three examples, corresponding to (a) $\phi(t_0) \gg H_0/2\pi$, (b)
$\phi(t_0) \simeq H_0/2\pi$, and (c) $ \phi(t_0) \ll H_0/2\pi$.

\begin{figure}
\epsfig{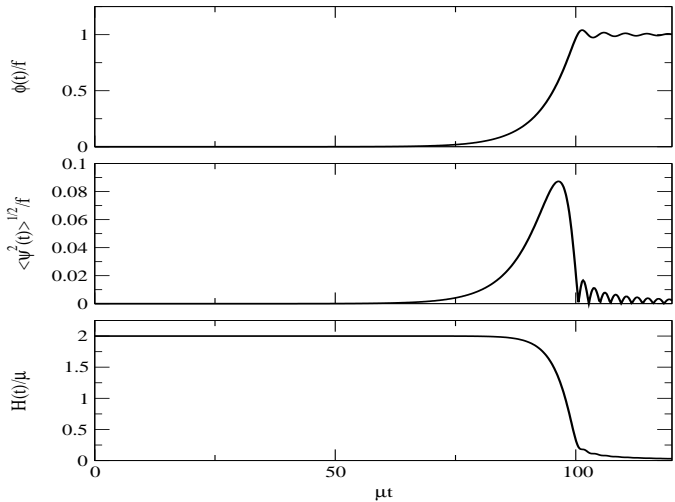}
\caption{The mean field $\phi(t)/f$, the fluctuation $<\psi^2(t)>^{1/2}/f$, 
  and the Hubble parameter $H(t)/\mu$ vs. $t$ for the $\lambda \Phi^4$ model
  with $\phi(t_0) = 5.0 H_0/2\pi$, $\dot{\phi}(t_0)=0$, $H_0=2\mu$, 
  $\lambda/8\pi^2=10^{-16}$, and $f \equiv \mu\sqrt{6/\lambda}$.}
\label{psh1}
\end{figure}

\begin{figure}
\epsfig{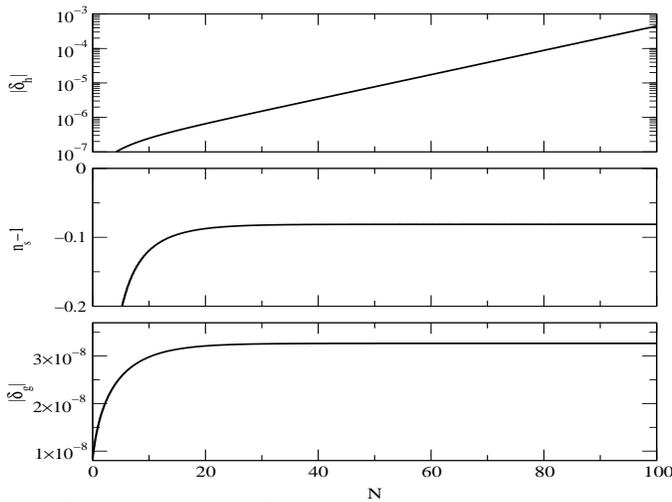}
\caption{The scalar amplitude $\delta_h$, the scalar tilt $n_s-1$,
  and the tensor amplitude $\delta_g$ vs. $N$ corresponding to the
  evolution of Fig.~\ref{psh1}.}
\label{dnd1}
\end{figure}

In the first of these, (a), shown in Fig.~\ref{psh1}, the evolution proceeds
exactly as would be expected from a purely classical analysis of the dynamics.
The two-point fluctuation $\langle \psi^2 \rangle$ remains small and does not
have a noticeable effect on the evolution of $\phi$, which in turn simply
follows the contour of the tree-level effective potential $V(\phi)$.  
As $\langle \psi^2 \rangle$ remains small, the expressions for the amplitude 
of primordial metric perturbations (\ref{deltaH}) reduces to the usual slow 
roll expression (\ref{slowrolldeltaH}) and we arrive at the standard results 
for $|\delta_h(k)|$, $n_s-1$, and $|\delta_g|$.  These quantities are 
depicted in Fig.~\ref{dnd1} as a function of $N$, the number of $e$-folds 
before the end of inflation at which the corresponding length scale crosses 
the horizon.

\begin{figure}
\epsfig{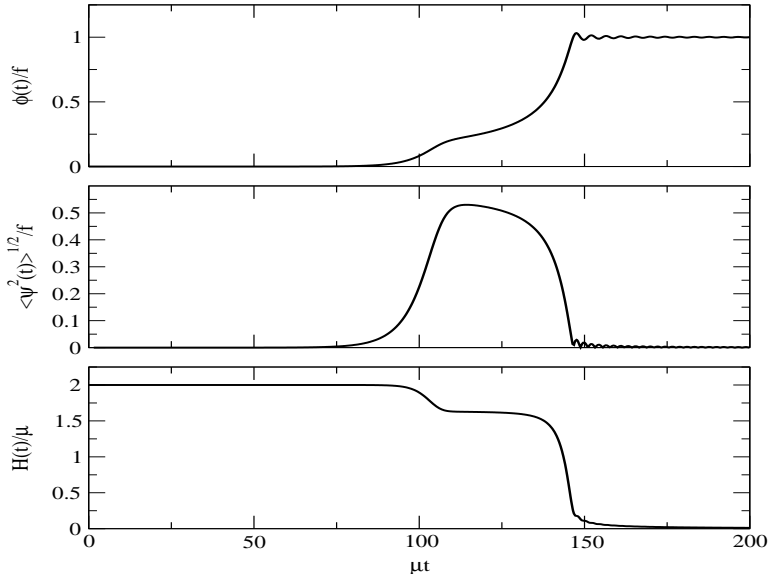}
\caption{The mean field $\phi(t)/f$, the fluctuation $<\psi^2(t)>^{1/2}/f$, 
  and the Hubble parameter $H(t)/\mu$ vs. $t$ for the $\lambda \Phi^4$ model
  with $\phi(t_0) = 0.4 H_0/2\pi$, $\dot{\phi}(t_0)=0$, $H_0=2\mu$,
  $\lambda/8\pi^2=10^{-16}$, and $f \equiv \mu\sqrt{6/\lambda}$.}
\label{psh2}
\end{figure}

Example (b), Fig.~\ref{psh2}, is an intermediate example for which 
the fluctuation becomes large
for a short time, inducing a spinodal phase
during the period of evolution for which the length scales
of relevance to CMB observations cross the horizon.  This example is of
further importance because it depicts the case in which the initial classical
value of the inflaton field is of the same order as its initial vacuum
fluctuations.  As is clearly shown, the growth of the
two-point fluctuation has a significant influence on the evolution of the mean
field $\phi$, resulting as well in a modified behavior for the expansion.

\begin{figure}
\epsfig{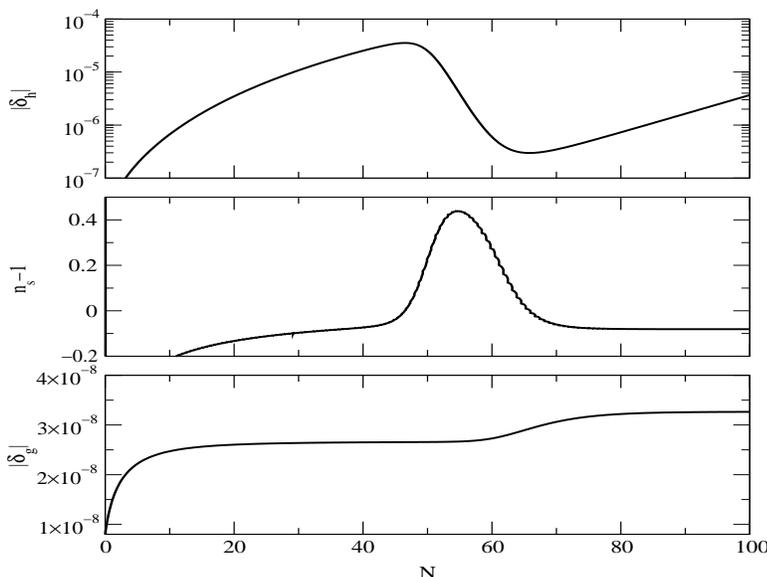}
\caption{The scalar amplitude $\delta_h$, the scalar tilt $n_s-1$,
  and the tensor amplitude $\delta_g$ vs. $N$ corresponding to the
  evolution of Fig.~\ref{psh2}.}
\label{dnd2}
\end{figure}

The result for the metric perturbations are provided in
Fig.~\ref{dnd2}, where we see some remarkable features.  First,
we notice that the amplitude of $|\delta_h(k)|$ is reduced
and that its shape is significantly changed by the spinodal dynamics.  The
significance of the shape is further emphasized by the scalar tilt $n_s-1$.  
Here, we see that for some scales of
observational relevance between $60$ and $50$ $e$-folds before the end of
inflation, the scalar tilt parameter $n_s-1$ is positive.  This corresponds to
a blue tilt in the power spectrum and was, until now, considered to be
unattainable in inflation models consisting of only a single scalar field.
Finally, the tensor amplitude clearly follows
the behavior of the expansion rate $H$ as expected.

\begin{figure}
\epsfig{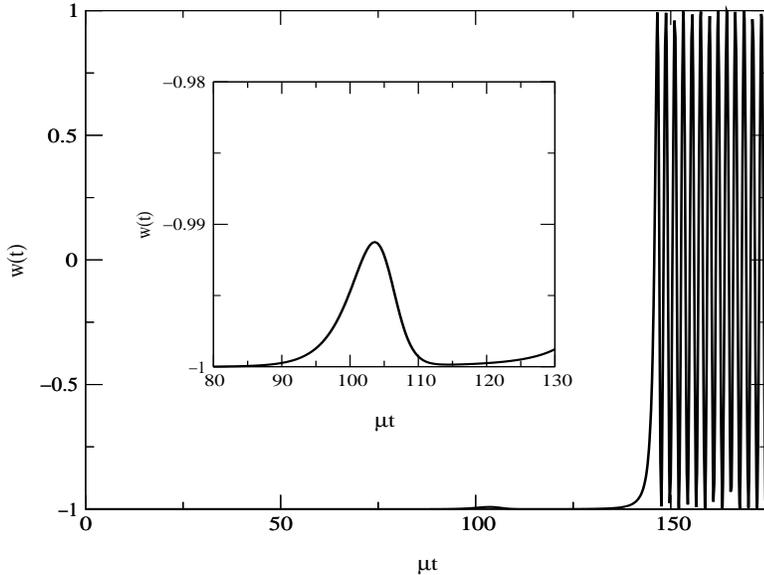}
\caption{The equation of state $w(t) \equiv p(t)/\varepsilon(t)$ vs. $t$ 
  for the parameters corresponding to the evolution of Fig.~\ref{psh2}.
  The expanded region shows the transition between the initial
  and spinodal phases of inflation.}
\label{eos2}
\end{figure}

To show why this is the case, we plot in Fig.~\ref{eos2} the equation of state
$w\equiv p/\rho$ as a function of time.  We see that in the transition region 
between the
initial and spinodal phases of inflation, there is little departure from the 
de Sitter equation of state $p = -\rho$ and that, in particular, $p < -\rho/3$
through to the end of the second inflationary stage.  This means that the
condition for accelerated expansion $\ddot{a} > 0$ is satisfied the whole
time, resulting in the simple behavior of the tensor amplitude.

Case (c), Fig.~\ref{psh3} shows the other extreme case when the initial 
classical value of the
field is significantly smaller than the effective quantum fluctuation.  Here,
$\langle \psi^2 \rangle$  reaches the spinodal while $\phi$
remains small, resulting in a very long spinodal phase
in which the mean field $\phi$ evolves slowly along the
spinodal line.  The results for the metric parameters are provided in
Fig.~\ref{dnd3}.  

\begin{figure}
\epsfig{file=psh3.eps,width=4.0in,height=3.0in}
\caption{The mean field $\phi(t)/f$, the fluctuation $<\psi^2(t)>^{1/2}/f$, 
  and the Hubble parameter $H(t)/\mu$ vs. $t$ for the $\lambda \Phi^4$ model
  with $\phi(t_0) = 0.05 H_0/2\pi$, $\dot{\phi}(t_0)=0$, $H_0=2\mu$, 
  $\lambda/8\pi^2=10^{-16}$, and $f \equiv \mu\sqrt{6/\lambda}$.}
\label{psh3}
\end{figure} 

\begin{figure}
\epsfig{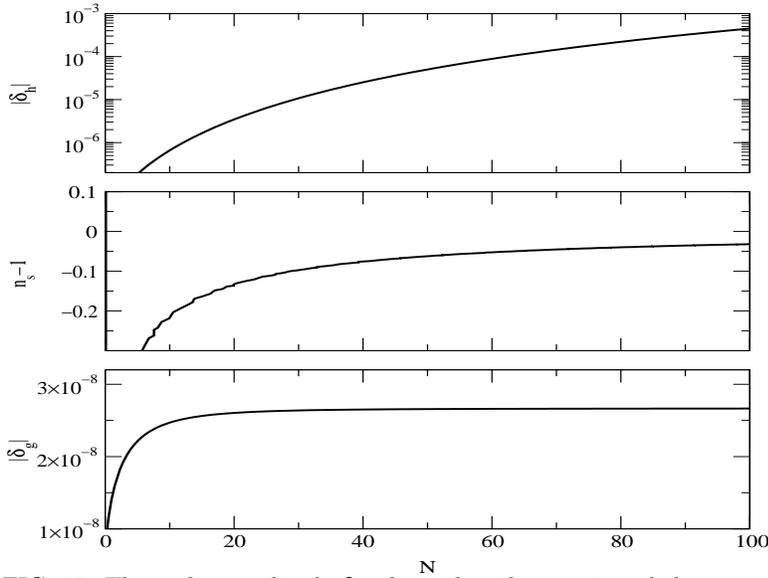}
\caption{The scalar amplitude $\delta_h$, the scalar tilt $n_s-1$,
  and the tensor amplitude $\delta_g$ vs. $N$ corresponding to the
  evolution of Fig.~\ref{psh3}.}
\label{dnd3}
\end{figure}

\begin{figure}
\epsfig{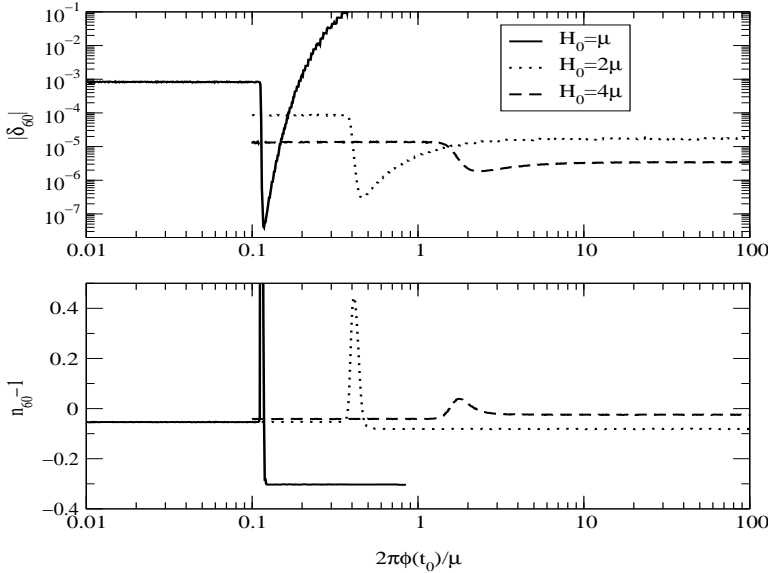}
\caption{The scalar amplitude $\delta_{60}$ and the scalar tilt 
  $n_{60}-1$ corresponding to the scale crossing the horizon $60$ $e$-folds
  before the end of inflation vs. $2\pi\phi(t_0)/\mu$ for 
  $\lambda \Phi^4$ new inflation using the classical two field model
  with $\dot{\phi}(t_0)=0$, 
  $\lambda/8\pi^2=10^{-16}$ and several values of $H_0$.}
\label{deltaphi}
\end{figure}

To add completeness to this picture, it is useful to plot the scalar amplitude
and tilt as a function of the log of the initial value $\phi(t_0)$ over the 
range of the differing regimes.  This is done for the particular scale 
crossing the horizon $60$ $e$-folds before the end of inflation 
(Fig.~\ref{deltaphi}).  This figure 
provides a nice summary of our primary results.

In the classical regime $\phi(t_0) \gg H_0/2\pi$, the results are independent
of initial conditions, but as $\phi(t_0)$ becomes of order $H_0/2\pi$ there is
a distinct transition regime in which the amplitude drops by up to a couple of
orders of magnitude (possibly alleviating the fine tuning problem of inflation
to a minor degree), while there is a spike for which the tilt becomes
positive.  While this region appears rather restricted in the parameter space
of initial values for $\phi$, we remind the reader that we plot here on a
{\em log} scale the ratio of $\phi(t_0)$ to $\mu/2\pi$ so that the region in
question effectively covers the entire range for which $\phi(t_0)$ and 
$H_0/2\pi$ are of the same order of magnitude. 

Finally, we see distinct regions for which there is a long spinodal phase.  
In this case, $\langle \psi^2 \rangle$ has a significant effect long
before the relevant scales cross the horizon.  The mean field
evolves slowly along the spinodal line to the spinodal point.  As the sum rule
(\ref{sumrule}) is in place during this phase, the evolution of $\phi$ in the
latter stages is always the same, and we therefore expect that $|\delta_{60}|$
and $n_{60}-1$ become constant throughout this regime, as is indeed the case.

It is worth mentioning how these results depend upon the other
parameters of the model, such as the magnitude of the quartic coupling
$\lambda$ and the ratio of the initial expansion rate to the mass scale
$H_0/\mu$.  

Increasing the value of $\lambda$ while keeping $H_0/\mu$ fixed only acts to
modify the value of the vacuum expectation value and the spinodal point to be 
smaller as both of
these quantities are proportional to $1/\sqrt{\lambda}$.  This simply reduces
the amount of time the fields evolve, but does not change any of the features
described above, with the exception that the scalar amplitude $|\delta_h(k)|$
scales, as usual, as $\sqrt{\lambda}$.

Modification of the ratio $H_0/\mu$ (keeping $\lambda$ fixed), on the other
hand has a more complex influence on the behavior.  The
results for the scalar amplitude and tilt as a function of initial conditions
for three values of $H_0/\mu$ are shown in Figs.~\ref{deltaphi}.
Here, we see that increasing the parameter has the effect of reducing the
amplitudes of the features in the transition region.  

Note that in the
case of $H_0/\mu=1$ the approximations we have used to compute the metric
perturbation spectrum break down in the transition region, where we see that
the scalar tilt spikes to a very large value, and therefore this computation
is unreliable for that small range of initial conditions.  However, as long as
the transition period does not occur within the last $60$ $e$-folds of
inflation, and in particular for the spinodal region for $\phi(t_0) \ll
H_0/2\pi$, the slow roll conditions are satisfied throughout the relevant
period of evolution such that the portions of this plot away from the
transition region are reliable.

\begin{figure}
\epsfig{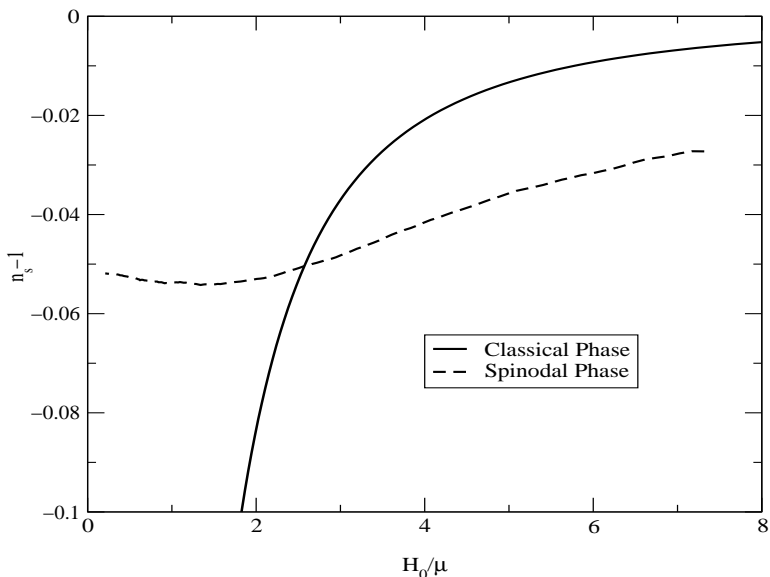}
\caption{The scalar tilt $n_{60}-1$ corresponding to the scale 
  crossing the horizon $60$ $e$-folds before the end of inflation 
  vs. $H_0/\mu$ for $\lambda \Phi^4$ new inflation using the classical two
  field model and spinodal initial
  conditions, ${\phi}(t_0) \ll H_0/2\pi$.}
\label{signature}
\end{figure}

Finally, we mention a somewhat unexpected feature of the results for 
the spinodal regime with small initial $\phi(t_0)$.  In Fig.~\ref{signature},
we plot the scalar tilt parameter $n_s-1$ as a function of the value of the
initial Hubble parameter $H_0/\mu$ and compare it to the result in the
classical regime for which $n_s-1 = \mu^2/3H_0^2$.  The result is remarkable.
Over a large range of initial expansion rates, there is very little change 
in the value of the scalar tilt.  It seems that the tilt due to a long
spinodal phase of inflation is relatively independent of the parameters
of the model, with an empirical prediction over the range of tested 
parameters of $0.94 < n_s < 0.98$.  While extension to even larger values of
$H_0/\mu$ may bring the upper limit toward $1.0$, it seems likely that the
lower limit will remain solid.  The consequence is that a future measurement 
which restricts the tilt to $n_s < 0.94$ would seem to rule out a spinodal 
phase of inflation.      

\subsection{natural inflation}
Due to its popularity, we examine one other spinodal inflation model in detail:
natural inflation.  The qualitative behavior is in many ways quite similar to
that of the $\lambda \Phi^4$ model described above, so we will focus on the
distinguishing features.  The primary difference occurs during the spinodal
phase.  In the $\lambda \Phi^4$ model, the spinodal condition of
Eq.~(\ref{sumrule}) simply results in the lowest order contribution to the
effective mass squared of the true zero mode $\phi$ being of order $-\lambda
\phi^2$, thus slowing the evolution of $\phi$. In natural inflation, however,
the spinodal condition for natural inflation yields an effective mass which
is {\it exponentially} suppressed by the growth of the fluctuations.  
Hence, the
evolution of the zero mode can come to practically a standstill.  The
net result is that the spinodal phase in natural inflation models is
significantly longer than a corresponding $\lambda \Phi^4$ model.

\begin{figure}
\epsfig{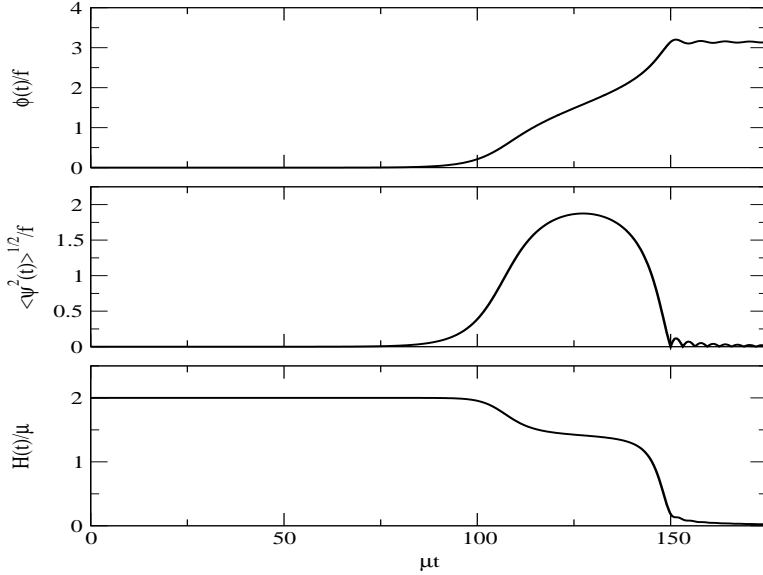}
\caption{The mean field $\phi(t)/f$, the fluctuation $<\psi^2(t)>^{1/2}/f$, 
  and the Hubble parameter $H(t)/\mu$ vs. $t$ for the natural inflation model
  with $\phi(t_0) = H_0/2\pi$, $\dot{\phi}(t_0)=0$, $H_0=2\mu$, and 
  $\Lambda^4/4\pi^2f^4=10^{-16}$.} 
\label{phisighub1}
\end{figure}

\begin{figure}
\epsfig{file=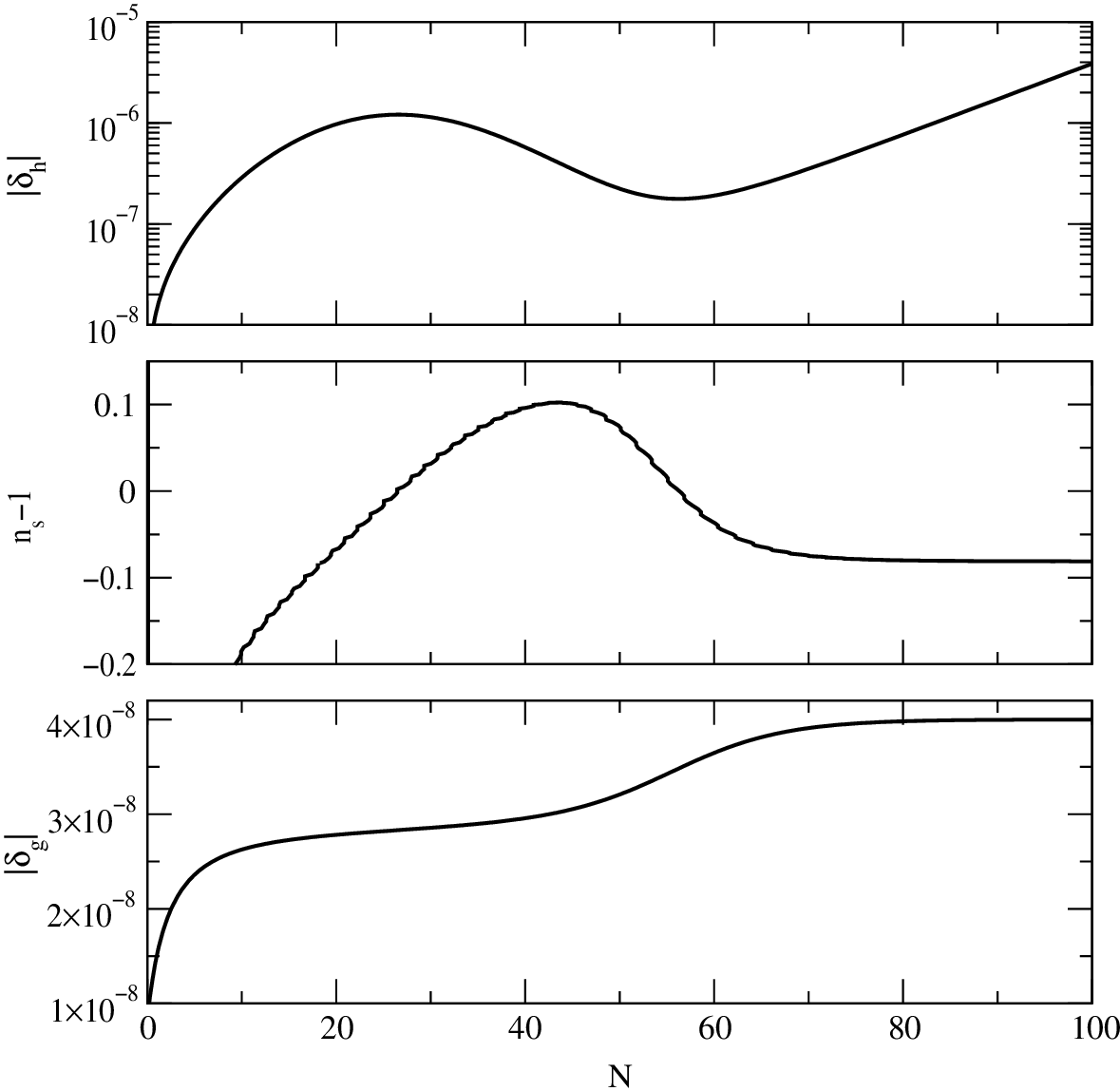,width=4.0in,height=3.0in}
\caption{The scalar amplitude $\delta_h$, the scalar tilt $n_s-1$,
  and the tensor amplitude $\delta_g$ vs. $N$ corresponding to the
  evolution of Fig.~\ref{phisighub1}.}
\label{dndnat1}
\end{figure}

We provide results for two examples corresponding to the intermediate case
with $\phi(t_0) \simeq H_0/2\pi$ shown in Figs.~\ref{phisighub1} and 
\ref{dndnat1} and the fully fluctuation dominated case of 
$\phi(t_0) \ll H_0/2\pi$ as depicted in Figs.~\ref{phisighub2} and 
\ref{dndnat2}.  We see the same features that appeared in the previous 
model, with the obvious difference that the spinodal phase of 
Fig.~\ref{phisighub2} is extremely long even for $\phi(t_0)$ not much
smaller than $H_0/2\pi$.

\begin{figure}
\epsfig{file=phisighub2.eps,width=4.0in,height=3.0in}
\caption{The mean field $\phi(t)/f$, the fluctuation $<\psi^2(t)>^{1/2}/f$, 
  and the Hubble parameter $H(t)/\mu$ vs. $t$ for the natural inflation model
  with $\phi(t_0) = 0.5 H_0/2\pi$, $\dot{\phi}(t_0)=0$, $H_0=2\mu$, and 
  $\Lambda^4/4\pi^2f^4=10^{-16}$.}
\label{phisighub2}
\end{figure} 

\begin{figure}
\epsfig{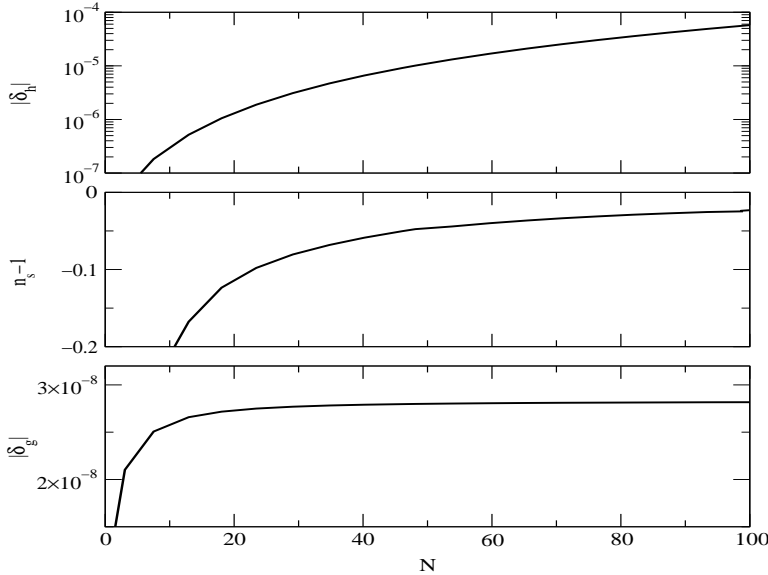}
\caption{The scalar amplitude $\delta_h$, the scalar tilt $n_s-1$,
  and the tensor amplitude $\delta_g$ vs. $N$ corresponding to the
  evolution of Fig.~\ref{phisighub2}.}
\label{dndnat2}
\end{figure}

\begin{figure}
\epsfig{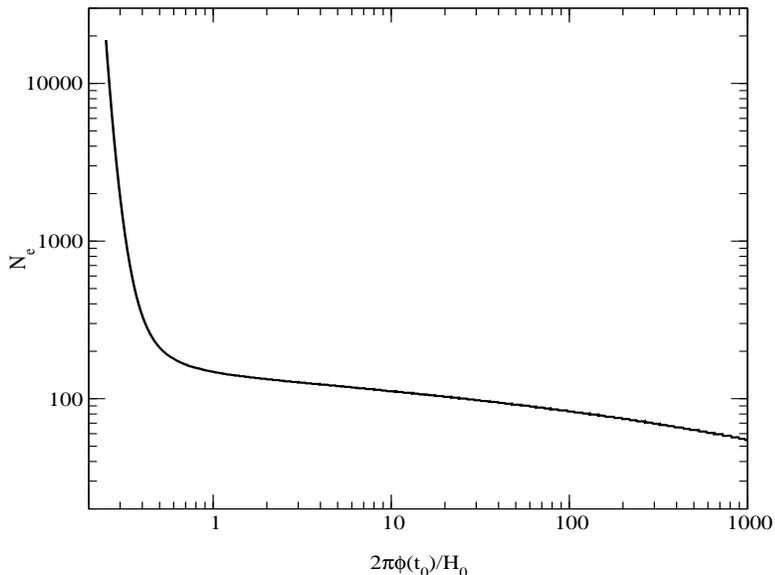}
\caption{The total number of $e$-folds of inflation $N$ vs. 
  $2\pi\phi(t_0)/H_0$ for natural inflation using the classical two
  field model 
  with $\dot{\phi}(t_0)=0$, $\Lambda^4/4\pi^2f^4=10^{-16}$ and 
  $H_0=2\Lambda^2/f$.}
\label{N}
\end{figure}

It is worth examining this point further by plotting the number of $e$-folds
of inflation as a function of the initial condition on $\phi$.  This is shown
in Fig.\ref{N} where we see a dramatic dependence which contrasts very
sharply with the logarithmic dependence of the same quantity on initial
conditions in the classical regime.

\begin{figure}
\epsfig{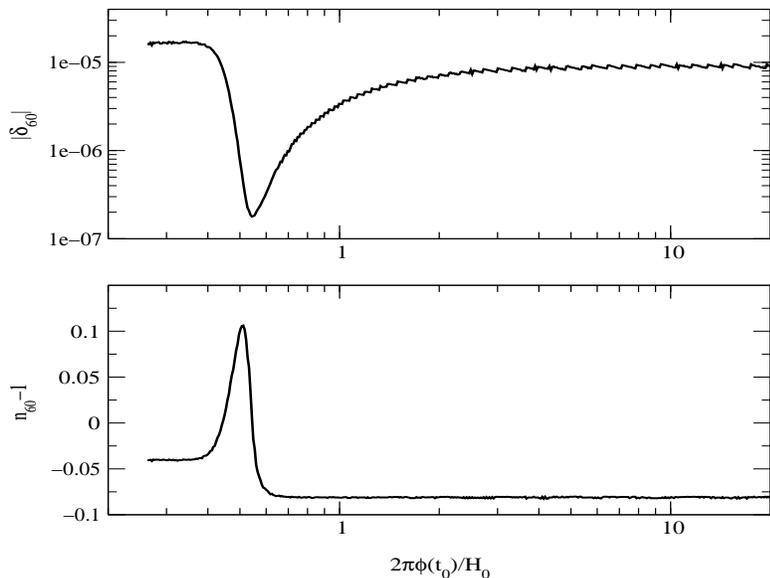}
\caption{The scalar amplitude $\delta_{60}$ and the scalar tilt 
  $n_{60}-1$ vs. $2\pi\phi(t_0)/H_0$ for natural inflation using the
  classical two field model 
  with $\dot{\phi}(t_0)=0$, $\Lambda^4/4\pi^2f^4=10^{-16}$ and 
  $H_0=2\Lambda^2/f$.}
\label{deltanat}
\end{figure}

Once again, it is enlightening to examine the dependence of the quantities
$|\delta_{60}|$ and $n_{60}-1$ as a function of the initial state.  These are
depicted in Figs.~\ref{deltanat} where we see again
the clear transition regime around $\phi(t_0) \sim H_0/2\pi$ separating the
classical and fluctuation dominated regimes.

\subsection{Example $C_l$ Spectrum}
\begin{figure}
\epsfig{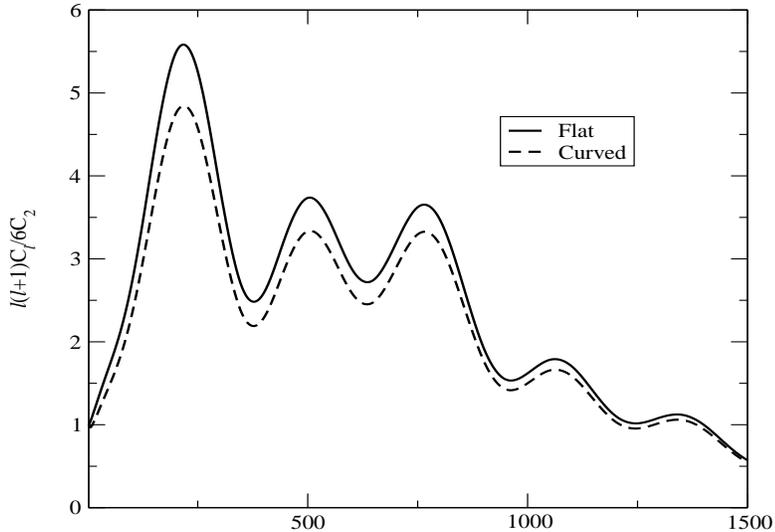}
\caption{The standard CDM $C_l$ spectrum of temperature anisotropies vs. $l$ 
  for a flat $n_s=1$ spectrum and the curved spectrum corresponding to the 
  evolution of Fig.~\ref{phisighub1}.}
\label{testcls}
\end{figure}

Finally, we present an example $C_l$ spectrum resulting from spinodal
inflation, shown in Fig.~\ref{testcls}. We plot the tilted $C_l$
spectrum corresponding to the natural inflation case of 
Fig.~\ref{phisighub1} as well as a standard flat $C_l$ spectrum for
comparison.  The `flat' spectrum was produced using the program 
CMBFAST version 2.4.1\cite{cmbfast} using the defaults of a standard 
Cold Dark Matter cosmology with a tiltless adiabatic power spectrum $n_s=1$, 
while the `curved' spectrum was reproduced from the flat one using the 
approximate relation (\ref{tiltedCls}).

The primary feature of interest is that while the spinodal
inflation spectrum is shifted downward for the modes $100 < l < 1000$, 
the spectrum approaches its flat counterpart for high $l$.  This is 
indicative of the shift of the spectrum from red to blue over the range
of observable scales.  

\section{Conclusions}
Whether one studies QCD, ferromagnetism, or ordinary gases, non-linear, long
wavelength effects are seen to have a dramatic impact on the properties of the
physical system.  Without close contact between the observation and theory of
these systems, there is good reason to doubt whether the respective phenomena 
of confinement, ferromagnetism, or the liquid-gas phase transition would be as
well understood as they are today.  But that is precisely the challenge before
us if we wish to understand the dynamics of phase transitions in the very early
stages of the universe.

If we expect to be able to proceed, then it will be necessary to return to
those systems that we believe we understand, say in condensed matter physics,
and examine the techniques, approximations, and concepts which have led to
consistent and accurate theoretical descriptions.  Looking in any statistical
physics text we immediately notice that there are a few general concepts and
techniques which have been found to be particularly rewarding for a variety of
different systems.  

One of these is mean field theory, where complex details of a system are
replaced by simple averages.  It is in many ways a very naive approach and
often provides quantitative results which are only roughly correct.  But the
real power of mean field theory lies not only in its quantitative predictions,
rather in the qualitative pictures it paints, allowing us, for example, to
explain the liquid-gas phase transition in terms of a system as simple as that
given by the van der Waals equation of state.

Another powerful concept is the convexity of the thermodynamical free energy
function for any equilibrium system and the simple rules for writing down such
a function provided by the Maxwell construction.  The story being told by
these concepts is that theoretical energy curves with concave portions
describe `unphysical' states, and that the dynamics of the corresponding
system will act to transform any such state into a physical state described by
a purely convex energy function.  

This report is an attempt to understand the system of an inflationary phase
transition in the context of these powerful concepts.  The initial state is
unphysical, described by a concave effective potential, the analog of the
equilibrium free energy, with quanta corresponding to imaginary mass states.
Such a system must decay into physical states, which we see from the
exponential growth of long wavelength fluctuations.  This decay continues
until a non-perturbative state is reached for which the corresponding quanta
are physical with a strictly real or zero mass.  

There are two possible ways to reach such a state.  The first is simply to
provide a significant bias to the state such that the mean value of the field
(i.e. the order parameter) reaches its true vacuum value where the field
quanta are well defined.

The second way, appropriate for systems with small order parameter, is to
allow the system to phase separate into domains for which the field has either
positive or negative value.  Rather than the order parameter moving along the
potential energy diagram (see Fig.~\ref{spinpot}), the field drops down into
the center of the diagram to the spinodal line.  At the spinodal line, the
field quanta are massless and physical, and this state of affairs may be
relatively long lasting, ending only when one phase becomes so much more
prevalent than the other that the system may relax into a definite vacuum
state throughout the system.

We therefore have arrived at a consistent physical picture of inflationary
phase transitions based upon mean field theory.  Already, at this naive level,
we have seen new phenomena which impact not only the evolution of the inflaton
field but also have important implications for the interpretation of recent 
and soon to come observational data.

As a final note, we emphasize that the techniques used here are really only a
first -- or, rather, a second -- approximation to a very complicated system of
interaction between an unstable scalar field and gravity at very high
energies.  There are a number of possible avenues which might be taken to
improve upon these results for the dynamics and, in particular, for the
predictions of observational quantities.

One direction is to move beyond mean field theory.  Gravitationally, this
means doing something more sophisticated than semi-classical gravity.  There
has been recent work in this regard within the context of perturbation theory
up to two loop order\cite{abramoetal}, and interest in this area has grown 
somewhat due to the
possibility of new phenomena in the context of preheating\cite{metricpreheat}.
However, the
non-perturbative dynamics of the scalar field studied here corresponds to
non-perturbative departures of the gravitational dynamics from that of a
purely classical background field so that techniques based upon perturbative
expansions are of little help.

In terms of the scalar field dynamics, one possible avenue that has received
attention is the $1/N$ expansion of the $O(N)$ vector model, which 
includes contributions beyond mean field theory at next to leading order 
(this approximation is also promising as it might 
be consistently implemented for gravity as well)\cite{larged}.  
Another alternative approach is to use 
variational methods to compute the dynamics of the
system\cite{gareth}, a technique which might also be combined with the $1/N$ 
expansion.
However, significant hurdles remain before either of these techniques will 
be implementable for interesting field theory problems.

The primary alternative to these semi-analytic approaches is to look to 
lattice simulations.  As the interesting
phenomena occur out of equilibrium, fully quantum simulations are ruled out 
and our only alternative is to examine real time simulations of a classical 
scalar coupled to classical gravity.  This approach has led, in particular, to
improvements in our understanding of preheating 
dynamics\cite{classpreheat,finelli}.  However, extending
these methods to interesting problems in inflationary dynamics will be a
challenge requiring a fully general relativistic lattice coupled to the
inflaton field, a way to properly deal with an exponentially changing scale,
and a method of introducing classical fluctuations consistent with the
classicalization of quantum, sub-horizon field modes as they cross the
horizon. 

The challenges are great, but we should be encouraged both by the improved
physical picture presented here and by the incredible fact that inflationary
theory is on the verge of becoming an observational science.

\acknowledgements
D.C. was supported by the Alexander von Humboldt Foundation.
R.H. was supported by DOE grant DE-FG02-91-ER40682.

\end{document}